\def\ni{\noindent}
\documentclass{aa}

\def\gsimeq{{_>\atop^{\sim}}}

\usepackage{rotating}
\usepackage{psfig}
\begin{document}

   \thesaurus{06     
              (03.11.1;  
               16.06.1;  
               19.06.1;  
               19.37.1;  
               19.53.1;  
               19.63.1)} 
   \title{A deep VLA survey at 6 cm in the Lockman Hole}

   \author{P. Ciliegi \inst{1}, G. Zamorani \inst{1,2}, G. Hasinger \inst{3,4},    
      I. Lehmann \inst{3,4}, G. Szokoly  \inst{3,4} , and G. Wilson  \inst{5,6} 
          }

   \offprints{P. Ciliegi; ciliegi@bo.astro.it}

   \institute {$^1$INAF : Osservatorio Astronomico di Bologna, Via Ranzani 1, 
I-40127 Bologna, Italy \\
   $^2$ Istituto di Radioastronomia, CNR, Via Gobetti 101, I-40129 Bologna, Italy \\ 
   $^3$Astrophysikalisches Institute Postdam, 
An der Sternwarte 16, D-14482, Postdam,  Germany \\
   $^4$ Max-Planck-Institut f\"ur extraterrestrische Physik
Giessenbachstra\ss e, D-85748 Garching, Germany \\ 
   $^5$ Brown University, Physics Department, Providence, RI 02912, USA \\
   $^6$ SIRTF Science Center, California Institute of Technology, 
Mail Code 220-6, 1200 East California Boulevard, Pasadena, CA 91125, USA } 


   \date{Received 07-05-02; accepted 18-11-02}

    \authorrunning{P. Ciliegi et al.}
    \titlerunning{A deep VLA survey at 6 cm in the Lockman Hole}

   \maketitle

   \begin{abstract}

We have obtained  a deep radio image with the Very Large Array
at 6 cm in the Lockman Hole.  The noise level in 
the central part of the field is $\sim11\mu$Jy.
From these data we have extracted a catalogue  of 63 radio sources 
with a maximum distance 
of 10 arcmin from the field center and with peak flux density 
greater than 4.5 times the local rms noise. 
The differential source counts  are in good agreement
with those obtained by other surveys. 
The analysis of the radio spectral index 
suggests a flattening of the average radio spectra  and an increase of 
the population of flat spectrum radio sources in the faintest flux bin. 
Cross correlation with the ROSAT/XMM X-ray sources list 
yields 13 reliable radio/X-ray associations, corresponding to 
$\sim$21\% of the radio sample. 
Most of these  
associations (8 out of 13) are classified as Type II AGN.

Using optical CCD (V and I) and   K$^{\prime}$ band data with 
approximate limits of 
V$\sim$25.5 mag, I$\sim$24.5 mag and 
K$^{\prime}\sim$20.2 mag, we found 
an optical identification for 58 of the 63 radio sources. This corresponds 
to an identification rate of $\sim$92\%, one of the highest percentages 
so far available. From the analysis of the colour-colour diagram and of the 
radio flux - optical magnitude diagram we have been able to select a 
subsample of radio sources whose optical counterparts are likely to be 
high redshift (z$>$0.5) early-type galaxies, hosting an Active Galactic 
Nucleus responsible of the radio activity. This class of objects, rather 
than a population of star-forming galaxies, appears to be the dominant 
population ($\gsimeq$50\%) in a 
5GHz selected sample with a flux limit as low as 
50 $\mu$Jy. 

We also find evidence that at these faint radio limits a large 
 fraction ($\sim$60\%) of the faintest
optical counterparts ($i.e.$ sources  in the magnitude range 
22.5$<$I$<$24.5 mag)  of the radio 
sources are Extremely Red Objects (EROs) with I-K$^{\prime}>$4 
 and combining our radio data with existing ISO data we conclude
 that these EROs sources are probably associated with 
high redshift, passively evolving elliptical galaxies.  
The six radio selected EROs  represent only  
$\sim$2\% of the optically selected EROs present in the field. 
If their luminosity is indeed a sign of AGN activity, the small
number of radio detections suggests  
that a small fraction of the EROS population contains an active nucleus.

   \end{abstract}

%

\section{Introduction}

Deep radio surveys reaching flux density of few $\mu$Jy have revealed 
a population of faint sources in excess with respect to the ``normal''
population of powerful radio galaxies. 
While the radio source counts above a few mJy are fully explained in terms
of a population of classical radio sources powered by active galactic nuclei
(AGN) and hosted by elliptical galaxies, below this flux  starts to appear
a population of star-forming galaxies similar to the nearby starburst
population  dominating the {\em Infrared Astronomical Satellite} (IRAS)
60 $\mu$m counts. The excellent correlation between radio and mid--infrared
emission for these objects (Condon, 1992; Yun et al., 2001) suggests
that indeed the radio emission in these galaxies is directly related
to the amount of star formation. For this reason, faint radio surveys
and identification of the optical counterparts can in principle be used,
together with optical and infrared surveys, to study the evolution of the star
formation history. This would require a clear understanding of the
flux level at which the star--forming galaxies do indeed become the dominant
population of faint radio sources. However, despite many dedicated efforts 
(see, for example, Benn et al., 1993; Hammer et al., 1995; 
Gruppioni et al., 1999;
Georgakakis et al., 2000; Prandoni et al., 2001) the relative fractions of the 
populations responsible of the sub-mJy radio counts (AGN, starburst, 
late and early type galaxies), are still far from being well established.

Optical spectroscopy for a complete sample of faint radio sources would
be the most direct way for a proper classification of the optical counterparts.
However, the very faint magnitudes of a significant fraction of the
objects associated to faint radio sources  makes this approach difficult
even for 8m--class telescopes. Alternatively, approximate classification
of the counterparts can be achieved using photometric (colours) and
radio (spectral index) data.

The purpose of this work is to shed some light into the nature of 
these sources by studying a new faint radio sample for which
we have analyzed the radio spectral properties and derived 
photometric optical identifications down to faint optical and K band
magnitudes. To further strengthen these goals, we have observed a region
of the sky at 6 cm centered in the Lockman Hole, where excellent data are 
already available at 20 cm (de Ruiter et al., 1997), in the far-infrared 
band (Fadda et al., 2002; Rodighiero et al. in preparation), 
in the near infrared and optical bands 
(Schmidt et al., 1998; Lehmann et al., 2000; Lehmann et al., 2001; 
Wilson et al., 2001) and in 
the X-ray band (Hasinger et al., 2001).  In Sect. 2 we 
give a general description of the radio observations, while 
the radio catalogue and the source counts are presented in Sect. 3. 
In Sect. 4 we present the associations of the 6 cm sources with 
the radio (20cm), near--infrared, optical, and X-ray sources.  Finally, 
Sect. 5 is devoted to the discussion of our results, while Sect. 6
summarizes our conclusions.


\section{The Radio Observations and data reduction}

The VLA observations were done in three runs of eleven 
hours each on January 16, 17 and 19, 1999 at  
4835 and 4885 MHz with a bandwidth of 50 MHz in C configuration. 
A total of seven pointings  in a hexagonal grid 
with a size of $\theta_{\rm FWHP}/\sqrt{2} \sim$6.4 arcmin 
(where $\theta_{\rm FWHP}$ is the full-width at half-power of the 
primary beam, 9 arcmin at 5 GHz) plus one at the center were 
observed for 4 hours each
around the ROSAT ultra-deep HRI field center  
RA(2000)= 10$^h52^m43^s$, DEC(2000)= 57$^{\circ}28^{\prime}
48^{\prime\prime}$. This  choice of pointing positions 
was adopted in order to obtain a reasonably uniform rms noise 
level in the inner part of the ROSAT ultra deep HRI field. 

All the data were analyzed using the NRAO AIPS reduction 
package. The data were calibrated using 3C 286 as primary 
flux density calibrator (assuming a flux density of 7.5103 Jy 
at 4835 MHz and 7.4617 Jy at 4885 MHz) and the source 1035+564 
as a phase and secondary amplitude calibrator. 

\begin{figure}
   \psfig{figure=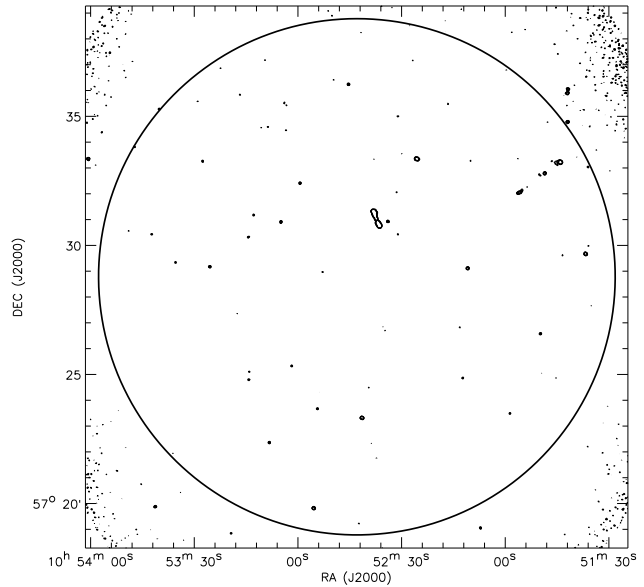,width=9cm}
      \caption[]{Contour plot of the 6 cm VLA image of the Lockman Hole. 
Contours are drawn starting from 4.5 times the local rms noise 
(see Fig.~\ref{F2}) with an 
increasing factor of $\sqrt{2}$. The circle of 10 arcmin radius shows 
the area used to extract the sources. }
\label{F1}
    \end{figure}

For each of the seven fields we constructed an image of 
1024$\times$1024 pixels, with a pixel-size of 1.0 arcsec.
Each observation was cleaned using the task IMAGR, 
using a restoring beam of 4$\times$4 arcsec. The rms noise 
levels in the cleaned (not primary beam corrected) 
images are uniform  and of the order of 11 $\mu$Jy. Finally, using the 
AIPS task LGEOM, HGEOM and LTESS, we have combined 
all the seven pointings, creating a single mosaic map of 
2048$\times$2048 pixels. Contour plot of the mosaic map is 
shown in Fig.~\ref{F1}.  

\begin{figure}
   \psfig{figure=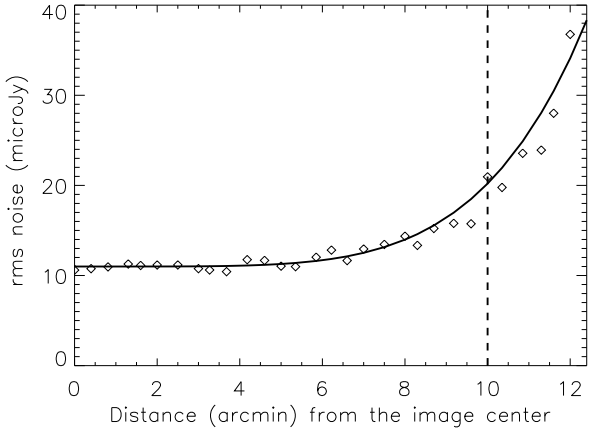,width=9cm}
      \caption[]{The rms noise (in $\mu$Jy) 
as a function of the distance from the image center. The solid line 
is the best fit function 
rms = 8.2$\times$ 10$^{-5} \times \theta_{off}^{5.05} $ + 11 $\mu$Jy. 
The dashed vertical line indicates the maximum distance from 
the center within which the sample described in this paper has been 
extracted.              }\label{F2}
    \end{figure}

As expected, the mosaic map has a 
regular noise distribution: a circular central region with a flat 
noise distribution, surrounded by an outer region where the noise 
increases for increasing distance from the center. No structures 
or irregularities were found in the rms noise map. The rms values 
as a function of the distance from the field center are well 
fitted by the function:  
rms($\theta_{off}$) = 8.2 $\times$ 10$^{-5} \times \theta_{off}^{5.05} $ + 11
$\mu$Jy where $\theta_{off}$ is the off-axis angle in arcmin. 
Measurements of the rms noise at various distances from the 
field center and the fitting function  are shown in Fig. ~\ref{F2}.  

Given the  rapid 
increase of the rms noise at large off-axis angles, we have limited the 
extraction of the sources to an area with $\theta_{off} \leq$10 arcmin, 
corresponding to 0.087 square degree.

\section{The Source Catalogue}

\subsection{Source detection}

The criterion we adopted for including a source in the catalogue 
is that its peak 
flux density, S$_{\rm P}$, is $\geq$4.5 times the average rms value at the 
off-axis angle of the source (see  Fig. ~\ref{F2}). 
In order to choose the threshold in  S$_{\rm P}$ we analyzed the map with the 
negative 
peaks using different threshold values (3.5, 4.0, 4.5, 5.0, 5.5 and 
6 respectively). 
The choice of 4.5 as threshold in S$_{\rm P}$ is a compromise 
between a  low  flux limit and the necessity to keep the 
number of spurious sources as low as possible. From the analysis 
of the negative peaks, we are confident that at most three or four 
spurious sources are present in the sample. 

The sources were extracted using the task SAD 
(Search And Destroy) which selects all 
the sources with peaks  brighter than a given level.
For each selected source the flux, the position and the size are
estimated using a least square Gaussian fit.  However the Gaussian 
fit may be unreliable both for faint and bright sources. In fact the 
noise map could influence the fit for faint sources 
(Condon, 1997), while the differences between the Gaussian beam and 
the real synthesized beam could in principle influence the 
goodness of a Gaussian 
fit even for sources with a large signal to noise ratio 
(Windhorst et al., 1984). 
Thus, we used 
SAD to extract all the sources whose peak flux S$_{\rm P}$ was greater
than  3 times 
the local rms value.  Subsequently, we derived
the peak flux for all the sources  using a second degree
interpolation (task MAXFIT). Only the sources with a MAXFIT 
peak flux density  $\geq$ 4.5 $\sigma$ were included in the final sample.
Hereafter we will use the MAXFIT peak flux density as the peak flux 
density S$_{\rm P}$ of the sources. 
For irregular and extended sources the total flux density
was determined by summing the values of all the pixels covering the source.
 
With this procedure we selected 63 radio sources (2 of which have multiple 
components for a total of 68 components) over a total area of 0.087 deg$^2$.

\begin{figure} 
\psfig{figure=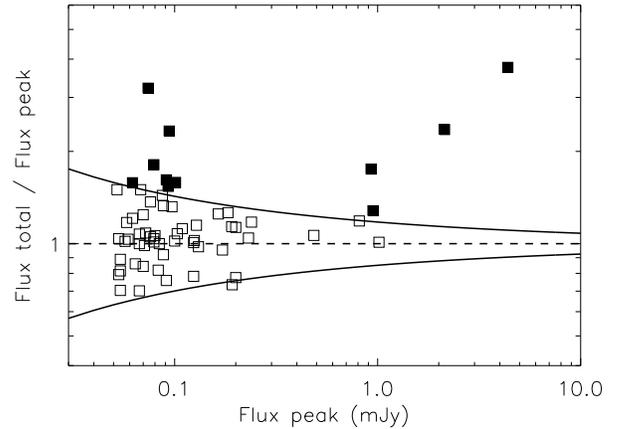,width=9cm} 
\caption[]{Ratio of the total flux, S$_{\rm T}$, to the peak flux, S$_{\rm P}$, 
as a function 
of the peak flux S$_{\rm P}$ for all the 6 cm radio sources in the Lockman Hole. 
The solid lines show the upper and lower envelopes of the flux ratio 
distribution  containing all the sources considered unresolved (open squares).
Filled squares show extended sources.}
\label{F3}
\end{figure}

In Fig.  ~\ref{F3} we  plot the ratio between the total 
(S$_{\rm T}$) and peak flux density (S$_{\rm P}$) 
as a function of the peak flux density for all the radio 
sources. Since the ratio of the total flux to the peak flux is a direct 
measure of the extension of a radio source, it can be used to discriminate 
between resolved or extended sources (i.e. larger than the beam) 
and unresolved sources. To select 
the extended sources, we have determined the lower envelope of the flux 
ratio distribution of  Fig. ~\ref{F3}  and we have mirrored it 
above the S$_{\rm T}$/S$_{\rm P}$=1 value (upper envelope of Fig.  ~\ref{F3}).  
We have considered extended the 12 sources laying above the upper envelope
that can be characterized by 
log (S$_{\rm T}$/S$_{\rm P}$) = $-$log (1-0.15 / S$_{\rm P}^{0.3}$).
However we should note that only 5 of the 12  sources classified 
as extended  lie significantly  above the upper 
envelope (see Fig.~\ref{F3})
and can be assumed as ``genuine'' extended sources. The other 7 sources 
just above the upper envelope might also  be  unresolved sources whose 
total flux has been overestimated due to the noise effect (see  
Windhorst et al. (1984) for a detailed discussion). 
Throughout the paper we have used the integrated flux for the 12 
formally extended sources 
and the peak flux for the unresolved sources in all the  calculations 
involving the radio flux.         

The catalogue 
with the 63 sources (68 components) is given in Table 1. 
For each source we report the name, the 
peak flux density S$_{\rm P}$ (and relative error $\sigma_{\rm S_{\rm P}}$)
in mJy, the total flux density S$_{\rm T}$ in mJy and error, 
the RA and DEC (J2000) and corresponding errors. 
Moreover, for resolved sources we 
also report the full width half maximum (FWHM) of the 
non deconvolved major and minor axes ($\theta_M$ and 
$\theta_m$ in arcsec) and  the position 
angle PA of the major axis (in degrees, measured east to north). 

All the errors in the source parameters were 
determined following the recipes given by Condon (1997). 
Also in the calculation of the flux error we used the 
integrated flux for  extended sources 
and the peak flux for unresolved sources. For the latter sources 
we used $\theta_M$ = $\theta_m$ = FWHM of synthesized beam = 4 arcsec. 

\begin{figure} 
\psfig{figure=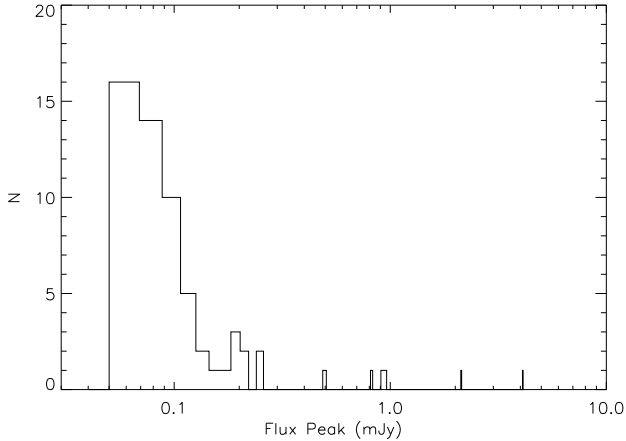,width=9cm} 
\caption[pos err]{Distribution of peak flux densities for 6 cm radio 
sources in the Lockman Hole}
\label{F4}
\end{figure}

 The different components of multiple sources are labeled ``A'', ``B'', 
etc., followed by a line labeled ``T'' in which flux and position 
for the total sources are given. 
Fig. ~\ref{F4} illustrates the distributions of peak flux 
densities for the 63 sources in the catalogue. 

\footnotesize

\begin{table*}
  \caption[]{The 4.5$\sigma$ radio sample}
  \label{catalog}
\begin{tabular}{lccccccccccr} 
&&&& \\ \hline
Name & S$_{\rm P}$ & $\sigma_{\rm S_{\rm P}}$ & S$_{\rm T}$ & $\sigma_{\rm S_{\rm T}}$ & R.A. (J2000) & DEC(J200) & $\sigma_{\alpha}$ & $\sigma_{\delta}$ & $\theta_M$ & $\theta_m$ & PA \\
     & (mJy) & (mJy) & (mJy) & (mJy) &  &  & ($\prime\prime$) & ($\prime\prime$) & ($\prime\prime$) & ($\prime\prime$) & deg  \\
&&&& \\ \hline       
\tiny LOCK\_6cm\_J105135+572739   &  0.079    &  0.017  &   0.079  &  0.017 &   10 51 35.42  &  57 27 39.6 & 0.59 & 0.51 &   0.0 &  0.0 &  0.0   \\
\tiny LOCK\_6cm\_J105136+572959   &  0.087    &  0.017  &   0.087  &  0.017 &   10 51 36.19  &  57 29 59.1 & 0.89 & 0.51 &   0.0 &  0.0 &  0.0   \\
\tiny LOCK\_6cm\_J105136+573302   &  0.200    &  0.020  &   0.200  &  0.020 &   10 51 36.24  &  57 33 ~2.6 & 0.38 & 0.45 &   0.0 &  0.0 &   0.0  \\
\tiny LOCK\_6cm\_J105137+572940   &  0.948    &  0.017  &   1.214  &  0.022 &   10 51 37.01  &  57 29 40.6 & 0.36 & 0.43 &   4.7 &  4.2 &  41.7  \\
\tiny LOCK\_6cm\_J105143+572938   &  0.074    &  0.014  &   0.238  &  0.046 &   10 51 43.77  &  57 29 38.7 & 0.59 & 0.92 &   9.1 &  6.0 &   8.9  \\
\tiny LOCK\_6cm\_J105143+573213   &  0.071    &  0.016  &   0.071  &  0.016 &   10 51 43.08  &  57 32 13.5 & 0.57 & 0.64 &   0.0 &  0.0 &   0.0  \\
\tiny LOCK\_6cm\_J105148+573248A  &  2.135    &  0.012  &   3.478  &  0.019 &   10 51 44.16  &  57 33 14.1 & 0.36 & 0.44 &  24.0 &  5.4 &   9.2  \\
\tiny LOCK\_6cm\_J105148+573248B  &  0.531    &  0.015  &   0.635  &  0.018 &   10 51 48.73  &  57 32 48.4 & 0.36 & 0.43 &   4.5 &  4.2 & -31.3  \\
\tiny LOCK\_6cm\_J105148+573248C  &  0.228    &  0.013  &   0.477  &  0.028 &   10 51 56.32  &  57 32 ~3.2 & 0.40 & 0.45 &   7.1 &  4.4 & -54.1  \\
\tiny LOCK\_6cm\_J105148+573248T  &  2.135    &  0.012  &   5.032  &  0.027 &   10 51 48.73  &  57 32 48.4 &  -   &  -   &    -  &   -  &     -  \\
\tiny LOCK\_6cm\_J105150+572635   &  0.192    &  0.013  &   0.192  &  0.013 &   10 51 50.12  &  57 26 35.5 & 0.38 & 0.45 &   0.0 &  0.0 &  0.0   \\
\tiny LOCK\_6cm\_J105150+573245   &  0.094    &  0.014  &   0.141  &  0.021 &   10 51 50.32  &  57 32 45.3 & 0.43 & 0.56 &   9.4 &  4.5 &  32.4  \\ 
\tiny LOCK\_6cm\_J105156+573322   &  0.067    &  0.014  &   0.067  &  0.014 &   10 51 56.47  &  57 33 22.4 & 0.45 & 0.48 &   0.0 &  0.0 &  0.0   \\
\tiny LOCK\_6cm\_J105158+572330   &  0.125    &  0.014  &   0.125  &  0.014 &   10 51 58.97  &  57 23 30.4 & 0.41 & 0.47 &   0.0 &  0.0 &  0.0   \\
\tiny LOCK\_6cm\_J105210+573317   &  0.062    &  0.012  &   0.062  &  0.012 &   10 52 10.14  &  57 33 17.5 & 0.41 & 0.76 &   0.0 &  0.0 &  0.0   \\
\tiny LOCK\_6cm\_J105211+572907   &  0.485    &  0.011  &   0.485  &  0.011 &   10 52 11.02  &  57 29 ~8.0 & 0.36 & 0.43 &   0.0 &  0.0 &  0.0   \\
\tiny LOCK\_6cm\_J105212+572453   &  0.124    &  0.012  &   0.124  &  0.012 &   10 52 12.49  &  57 24 53.0 & 0.39 & 0.45 &   0.0 &  0.0 &  0.0   \\
\tiny LOCK\_6cm\_J105213+572650   &  0.070    &  0.011  &   0.070  &  0.011 &   10 52 13.33  &  57 26 50.5 & 0.46 & 0.49 &   0.0 &  0.0 &  0.0   \\
\tiny LOCK\_6cm\_J105216+573529   &  0.076    &  0.013  &   0.076  &  0.013 &   10 52 16.61  &  57 35 30.0 & 0.48 & 0.53 &   0.0 &  0.0 &  0.0   \\
\tiny LOCK\_6cm\_J105225+573322   &  0.928    &  0.012  &   1.625  &  0.021 &   10 52 25.63  &  57 33 22.5 & 0.36 & 0.43 &   6.1 &  4.5 &  71.4  \\
\tiny LOCK\_6cm\_J105226+573711   &  0.080    &  0.015  &   0.080  &  0.015 &   10 52 26.79  &  57 37 11.1 & 0.65 & 0.56 &   0.0 &  0.0 &  0.0   \\
\tiny LOCK\_6cm\_J105227+573832   &  0.091    &  0.020  &   0.091  &  0.020 &   10 52 27.69  &  57 38 32.1 & 0.43 & 0.56 &   0.0 &  0.0 &  0.0   \\
\tiny LOCK\_6cm\_J105229+573334   &  0.054    &  0.011  &   0.054  &  0.011 &   10 52 29.99  &  57 33 34.6 & 0.47 & 0.54 &   0.0 &  0.0 &  0.0   \\
\tiny LOCK\_6cm\_J105231+573027   &  0.068    &  0.011  &   0.068  &  0.011 &   10 52 31.13  &  57 30 27.1 & 0.48 & 0.63 &   0.0 &  0.0 &  0.0   \\
\tiny LOCK\_6cm\_J105231+573204   &  0.067    &  0.011  &   0.067  &  0.011 &   10 52 31.48  &  57 32 ~5.0 & 0.53 & 0.51 &   0.0 &  0.0 &  0.0   \\
\tiny LOCK\_6cm\_J105231+573501   &  0.076    &  0.012  &   0.076  &  0.012 &   10 52 31.08  &  57 35 ~1.4 & 0.63 & 0.48 &   0.0 &  0.0 &  0.0   \\
\tiny LOCK\_6cm\_J105233+573057   &  0.231    &  0.011  &   0.231  &  0.011 &   10 52 33.99  &  57 30 57.1 & 0.37 & 0.44 &   0.0 &  0.0 &  0.0   \\
\tiny LOCK\_6cm\_J105234+572643   &  0.054    &  0.011  &   0.054  &  0.011 &   10 52 34.84  &  57 26 43.6 & 0.43 & 0.54 &   0.0 &  0.0 &  0.0   \\
\tiny LOCK\_6cm\_J105235+572640   &  0.054    &  0.011  &   0.054  &  0.011 &   10 52 35.19  &  57 26 40.2 & 0.77 & 0.59 &   0.0 &  0.0 &  0.0   \\
\tiny LOCK\_6cm\_J105237+572147   &  0.059    &  0.013  &   0.059  &  0.013 &   10 52 37.34  &  57 21 47.1 & 0.60 & 0.52 &   0.0 &  0.0 &  0.0   \\
\tiny LOCK\_6cm\_J105237+573104A  &  4.376    &  0.012  &   5.823  &  0.015 &   10 52 36.40  &  57 30 46.9 & 0.36 & 0.43 &   4.6 &  4.4 & -41.8  \\
\tiny LOCK\_6cm\_J105237+573104B  &  0.750    &  0.012  &   2.863  &  0.045 &   10 52 37.39  &  57 31 04.1 & 0.36 & 0.43 &   6.9 &  5.7 &  21.2  \\
\tiny LOCK\_6cm\_J105237+573104C  &  1.101    &  0.011  &   3.384  &  0.034 &   10 52 37.77  &  57 31 12.3 & 0.36 & 0.43 &  10.0 &  5.8 &  12.3  \\
\tiny LOCK\_6cm\_J105237+573104D  &  2.172    &  0.012  &   3.770  &  0.020 &   10 52 38.17  &  57 31 20.3 & 0.36 & 0.43 &   4.9 &  4.2 & -89.1  \\
\tiny LOCK\_6cm\_J105237+573104T  &  4.376    &  0.012  &  16.452  &  0.044 &   10 52 37.05  &  57 30 58.8 &  -   &  -   &    -  &   -  &  -     \\
\tiny LOCK\_6cm\_J105238+572221   &  0.057    &  0.012  &   0.057  &  0.012 &   10 52 38.78  &  57 22 21.4 & 0.57 & 0.59 &  0.0  &  0.0 &  0.0   \\
\tiny LOCK\_6cm\_J105238+573321   &  0.053    &  0.011  &   0.053  &  0.011 &   10 52 38.00  &  57 33 21.4 & 0.50 & 0.51 &  0.0  &  0.0 &  0.0   \\
\tiny LOCK\_6cm\_J105239+572430   &  0.058    &  0.011  &   0.058  &  0.011 &   10 52 39.54  &  57 24 30.7 & 0.61 & 0.53 &  0.0  &  0.0 &  0.0   \\
\tiny LOCK\_6cm\_J105241+572320   &  0.812    &  0.011  &   0.812  &  0.011 &   10 52 41.45  &  57 23 20.6 & 0.36 & 0.43 &  0.0  &  0.0 &  0.0   \\
\tiny LOCK\_6cm\_J105242+571915   &  0.101    &  0.017  &   0.160  &  0.027 &   10 52 42.47  &  57 19 15.8 & 0.66 & 0.81 &  8.5  &  4.3 &  38.4  \\
\tiny LOCK\_6cm\_J105245+573615   &  0.239    &  0.013  &   0.239  &  0.013 &   10 52 45.35  &  57 36 15.9 & 0.38 & 0.44 &  0.0  &  0.0 &  0.0   \\
\tiny LOCK\_6cm\_J105249+573626   &  0.070    &  0.013  &   0.070  &  0.013 &   10 52 49.73  &  57 36 26.6 & 0.66 & 0.59 &  0.0  &  0.0 &  0.0   \\
\tiny LOCK\_6cm\_J105252+572859   &  0.062    &  0.011  &   0.098  &  0.018 &   10 52 52.79  &  57 28 59.7 & 0.71 & 0.50 &  6.5  &  4.8 &  84.6  \\
\tiny LOCK\_6cm\_J105254+572341   &  0.131    &  0.011  &   0.131  &  0.011 &   10 52 54.27  &  57 23 41.7 & 0.39 & 0.45 &  0.0  &  0.0 & 0.0    \\
\tiny LOCK\_6cm\_J105255+571950   &  1.015    &  0.017  &   1.015  &  0.017 &   10 52 55.32  &  57 19 50.6 & 0.36 & 0.43 &  0.0  &  0.0 & 0.0    \\
\tiny LOCK\_6cm\_J105259+573226   &  0.164    &  0.011  &   0.164  &  0.011 &   10 52 59.34  &  57 32 26.1 & 0.38 & 0.46 &  0.0  &  0.0 &  0.0   \\
\tiny LOCK\_6cm\_J105301+572521   &  0.103    &  0.011  &   0.103  &  0.011 &   10 53 ~1.69  &  57 25 21.2 & 0.40 & 0.47 &  0.0  &  0.0 &   0.0  \\
\tiny LOCK\_6cm\_J105301+573333   &  0.052    &  0.011  &   0.052  &  0.011 &   10 53 ~1.48  &  57 33 34.0 & 1.01 & 0.51 &  0.0  &  0.0 &  0.0   \\
\tiny LOCK\_6cm\_J105303+573429   &  0.067    &  0.012  &   0.067  &  0.012 &   10 53 ~3.38  &  57 34 29.1 & 0.50 & 0.51 &  0.0  &  0.0 &  0.0   \\
\tiny LOCK\_6cm\_J105303+573526   &  0.072    &  0.013  &   0.072  &  0.013 &   10 53 ~3.28  &  57 35 26.9 & 0.50 & 0.53 &  0.0  &  0.0 &  0.0   \\
\tiny LOCK\_6cm\_J105303+573532   &  0.079    &  0.014  &   0.143  &  0.025 &   10 53 ~3.91  &  57 35 32.2 & 0.48 & 0.70 &  6.6  &  4.6 &  -7.7  \\
\tiny LOCK\_6cm\_J105304+573055   &  0.183    &  0.011  &   0.183  &  0.011 &   10 53 ~4.83  &  57 30 55.9 & 0.37 & 0.45 &  0.0  &  0.0 &  0.0   \\
\tiny LOCK\_6cm\_J105308+572223   &  0.191    &  0.013  &   0.191  &  0.013 &   10 53 ~8.09  &  57 22 23.0 & 0.38 & 0.45 &  0.0  &  0.0 &  0.0   \\
\hline   
\end{tabular}                                    

\end{table*}

\setcounter{table}{0}
\begin{table*}
  \caption{(continued)}
  \label{catalog}
\begin{tabular}{lccccccccccr} 
&&&& \\ \hline
Name & S$_{\rm P}$ & $\sigma_{\rm S_{\rm P}}$ & S$_{\rm T}$ & $\sigma_{\rm S_{\rm T}}$ & R.A. (J2000) & DEC(J200) & $\sigma_{\alpha}$ & $\sigma_{\delta}$ & $\theta_M$ & $\theta_m$ & PA \\ 

     & (mJy) & (mJy) & (mJy) & (mJy) &  &  & ($\prime\prime$) & ($\prime\prime$) & ($\prime\prime$) & ($\prime\prime$) & deg  \\

&&&& \\ \hline       
\tiny LOCK\_6cm\_J105308+573436   &  0.083   & 0.012   & 0.083   & 0.012  &   10 53 ~8.65  &  57 34 36.7 & 0.44 & 0.46  &  0.0  &  0.0 &  0.0   \\
\tiny LOCK\_6cm\_J105310+573434   &  0.064   & 0.012   & 0.064   & 0.012  &   10 53 10.57  &  57 34 34.8 & 0.47 & 0.53  &  0.0  &  0.0 &  0.0   \\
\tiny LOCK\_6cm\_J105312+573111   &  0.097   & 0.011   & 0.097   & 0.011  &   10 53 12.70  &  57 31 11.9 & 0.42 & 0.52  &  0.0  &  0.0 &  0.0   \\
\tiny LOCK\_6cm\_J105313+572507   &  0.079   & 0.011   & 0.079   & 0.011  &   10 53 13.89  &  57 25 ~7.4 & 0.53 & 0.46  &  0.0  &  0.0 &  0.0   \\
\tiny LOCK\_6cm\_J105314+572448   &  0.091   & 0.012   & 0.147   & 0.020  &   10 53 14.01  &  57 24 48.8 & 0.47 & 0.52  &  5.4  &  4.7 &  49.9  \\
\tiny LOCK\_6cm\_J105314+573020   &  0.093   & 0.012   & 0.143   & 0.018  &   10 53 14.20  &  57 30 20.6 & 0.51 & 0.48  &  6.1  &  4.0 & -67.4  \\
\tiny LOCK\_6cm\_J105316+573551   &  0.088   & 0.015   & 0.088   & 0.015  &   10 53 16.74  &  57 35 51.7 & 0.54 & 0.58  &  0.0  &  0.0 &  0.0   \\
\tiny LOCK\_6cm\_J105317+572722   &  0.053   & 0.011   & 0.053   & 0.011  &   10 53 17.39  &  57 27 22.4 & 0.65 & 0.51  &  0.0  &  0.0 &  0.0   \\
\tiny LOCK\_6cm\_J105322+573652   &  0.100   & 0.019   & 0.100   & 0.019  &   10 53 22.37  &  57 36 52.9 & 0.49 & 0.56  &  0.0  &  0.0 &  0.0   \\
\tiny LOCK\_6cm\_J105325+572911   &  0.200   & 0.012   & 0.200   & 0.012  &   10 53 25.30  &  57 29 11.5 & 0.38 & 0.44  &  0.0  &  0.0 &  0.0   \\
\tiny LOCK\_6cm\_J105327+573316   &  0.128   & 0.013   & 0.128   & 0.013  &   10 53 27.46  &  57 33 16.6 & 0.39 & 0.50  &  0.0  &  0.0 &  0.0   \\
\tiny LOCK\_6cm\_J105328+573535   &  0.084   & 0.017   & 0.084   & 0.017  &   10 53 28.92  &  57 35 35.8 & 0.56 & 0.52  &  0.0  &  0.0 &  0.0   \\
\tiny LOCK\_6cm\_J105335+572157   &  0.093   & 0.021   & 0.144   & 0.032  &   10 53 35.78  &  57 21 57.3 & 0.59 & 0.80  &  6.7  &  4.7 & -31.4  \\
\tiny LOCK\_6cm\_J105335+572921   &  0.109   & 0.013   & 0.109   & 0.013  &   10 53 35.18  &  57 29 21.3 & 0.41 & 0.49  &  0.0  &  0.0 &  0.0   \\
\tiny LOCK\_6cm\_J105342+573026   &  0.124   & 0.014   & 0.124   & 0.014  &   10 53 42.08  &  57 30 26.5 & 0.43 & 0.46  &  0.0  &  0.0 &  0.0   \\
\tiny LOCK\_6cm\_J105347+573349   &  0.172   & 0.020   & 0.172   & 0.020  &   10 53 47.14  &  57 33 49.5 & 0.41 & 0.47  &  0.0  &  0.0 &  0.0   \\
\tiny LOCK\_6cm\_J105348+573033   &  0.088   & 0.016   & 0.088   & 0.016  &   10 53 48.73  &  57 30 33.8 & 0.46 & 0.52  &  0.0  &  0.0 &  0.0   \\

\hline  
\end{tabular}                                    

\end{table*}

\normalsize

\subsection{Position Errors}

The projection of the major and minor axis errors onto the right ascension and 
declination axes produces the total rms position errors given by Condon et al. (1998)

\begin{equation}
\sigma^2_{\alpha} = \varepsilon^2_{\alpha} + \sigma^2_{x_0} \sin^2(PA) + 
\sigma^2_{y_0} \cos^2(PA) \label{eq:ra_rms}
\end{equation}
\begin{equation}
\sigma^2_{\delta} = \varepsilon^2_{\delta} + \sigma^2_{x_0} \cos^2(PA) + 
\sigma^2_{y_0} \sin^2(PA) \label{eq:dec_rms}
\end{equation}

\ni where ($\varepsilon_{\alpha}, \varepsilon_{\delta}$) are the 
``calibration'' 
errors,  $\sigma_{x_0}^2$ = $\theta_M^2/(4 ln 2)\rho^2$, 
$\sigma_{y_0}^2$= $\theta_m^2/(4 ln 2)\rho^2$
with $\rho$ the effective signal to noise ratio 
as defined in  Condon (1997). 

\begin{figure}
\psfig{figure=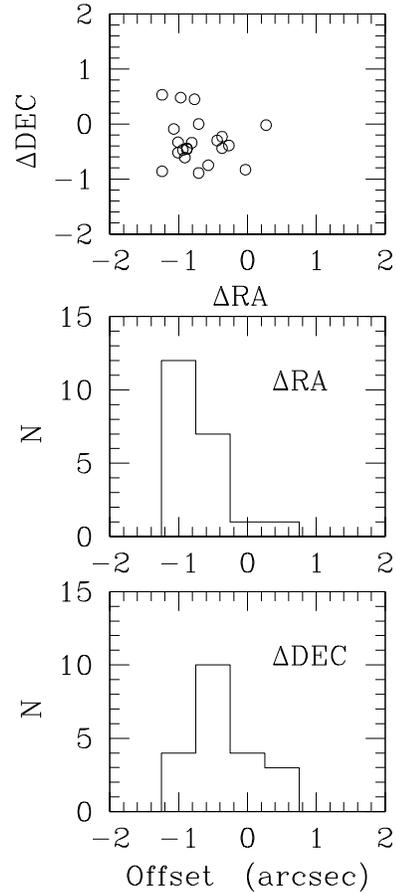,width=12cm}
\caption{Position offset for strong point sources (20 sources in 
our survey detected with S$_{\rm P}>6\sigma$ and  
with a point like-optical counterpart in the V band CCD)}
\label{F5}
\end{figure}

The mean image offset $<\Delta\alpha>$, $<\Delta\delta>$ 
(defined as radio minus optical position) and rms calibration
uncertainties $\varepsilon^2_{\alpha}$ and $\varepsilon^2_{\delta}$ are best
determined by comparison with  accurate positions of sources strong enough 
that the noise plus confusion terms are much smaller than the calibration
terms. We used the 6 cm positions of 20 strong compact sources (detected with 
S$_{\rm P}>6\sigma$) identified with point like optical counterparts 
in the V band CCD 
(i.e. we excluded the sources that have an extended or a multiple optical 
counterparts). Their offsets $\Delta\alpha$ and  
$\Delta\delta$ are shown in Fig. ~\ref{F5}. The mean 
offsets are $<\Delta\alpha>=-0.71\pm0.09^{\prime\prime}$ and 
$<\Delta\delta>=-0.31\pm0.09^{\prime\prime}$. These offsets
have not been applied to the radio positions 
reported in Table 1, but they should be 
removed (adding 0.71$^{\prime\prime}$ in RA and  
0.31$^{\prime\prime}$ in DEC at the radio position)   
to obtain the radio position  in the same reference frame as the optical CCD.

\begin{figure} 
\psfig{figure=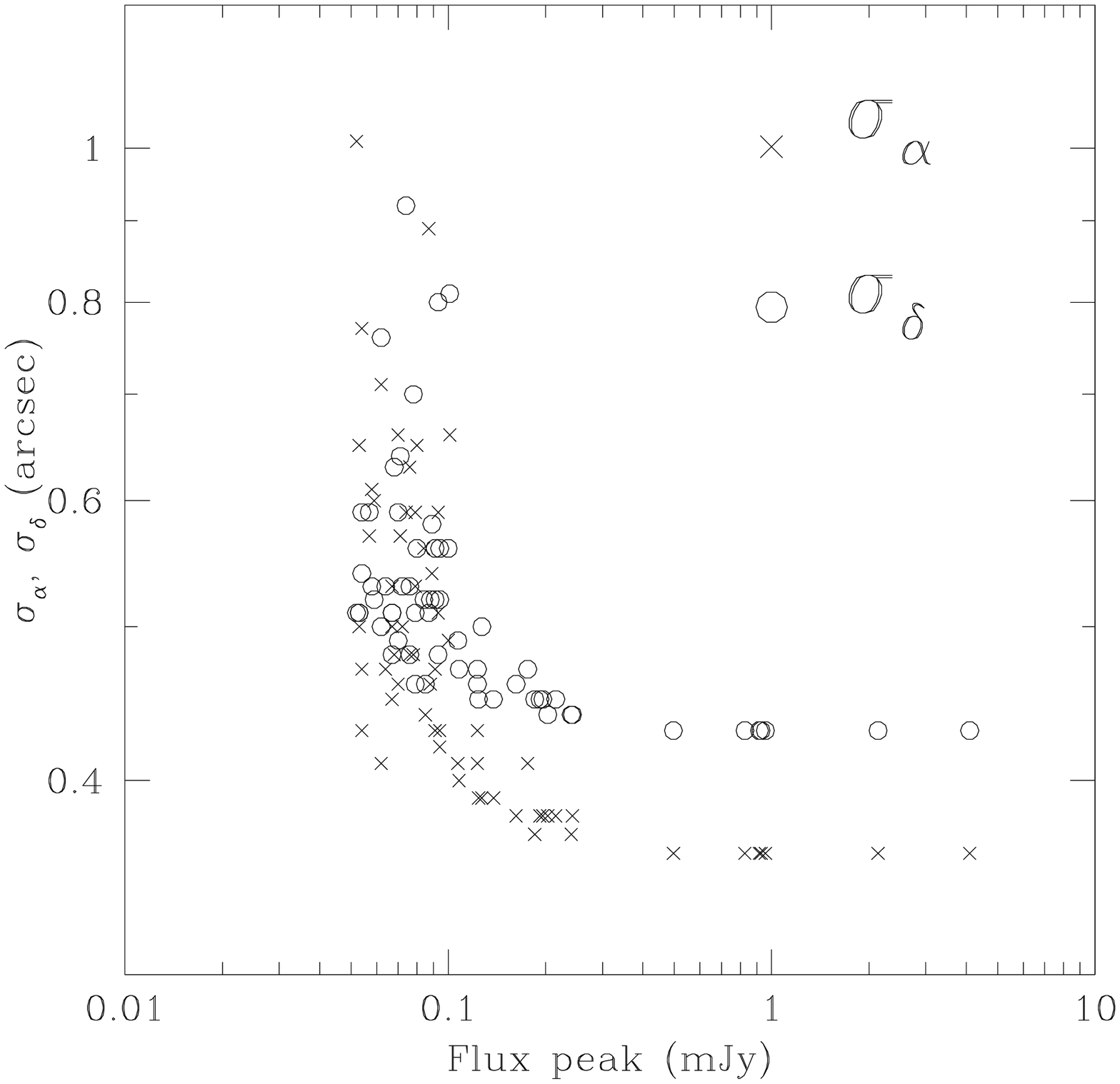,width=8cm} 
\caption[pos err]{The  position uncertainties $\sigma_{\alpha}$ 
and $\sigma_{\delta}$  
for all the sources  in the 6 cm catalogue as a function of 
the peak flux density $S_{\rm P}$.} 
\label{F6}
\end{figure}

From the distributions of $\Delta\alpha$ and  $\Delta\delta$ shown in 
Fig. ~\ref{F5} we have estimated the rms calibration errors: 
$\varepsilon_{\alpha}=0.36^{\prime\prime}$ and 
$\varepsilon_{\delta}=0.43^{\prime\prime}$. Since the 20 objects used for this 
derivation are scattered over the entire field of view, the 
small values of $\varepsilon_{\alpha}$ and $\varepsilon_{\delta}$ show that
our mosaic maps are not affected by relevant geometric distortions induced 
by the 
approximation  of a finite portion of the spherical sky 
with a bi-dimensional plane (see Perley, 1989; Condon 
et al., 1998).  Using these calibration errors, positional uncertainties 
for all  sources have been calculated according 
to Eqs. (1) and (2). 
The rms position uncertainties ($\sigma_{\alpha}$, $\sigma_{\delta}$) of
all the sources in our catalogue are reported in Table 1 (see Cols. 8 and 9) 
and plotted as a function of peak flux in 
Fig. \ref{F6}.

\subsection{Survey Completeness and Source Counts}

\begin{figure} 
\psfig{figure=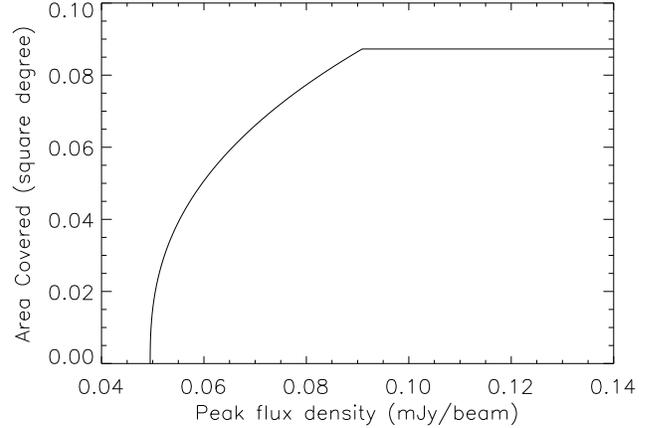,width=9cm} 
\caption[]{Areal coverage of the Lockman Hole at 6 cm  represented by the
  solid angle over which a source with peak flux $S_{\rm P}$ can be detected.}
\label{F7}
\end{figure}

The sample of 63 sources listed in Table 1 has been used to construct
the source counts distributions. The rms noise as a function of distance 
from the center (Fig.~\ref{F2}) was used to obtain 
the detectability area as 
a function of flux density. In Fig.~\ref{F7} the solid angle 
over which a source with a peak flux density S$_{\rm P}$ can be detected  
is plotted as a function of flux density.  

The two sources with multiple components 
have been treated as a single radio source. In computing the counts 
we have used the integrated flux for extended sources and the peak flux 
for unresolved sources (see Sect. 3.1 for the definition of 
resolved and unresolved sources).  

\subsubsection{Completeness} 

Before discussing the source counts (see next Section) we describe here 
the simulations that we performed  
in order to estimate the combined effect of noise, source extraction 
technique and resolution bias on the completeness of our sample.  
We constructed a simulated sample of radio sources down to a flux level 
of 0.025 mJy with  a density  
in the sky given by  N($>$S) = 0.42 $\times$ (S/30)$^{-1.18}$  
(where N($>$S) is the number of sources (arcmin)$^{-2}$ 
with total flux greater than S in $\mu$Jy, Fomalont et al., 1991)
and an appropriate angular size distribution.
Using the relation reported by Windhorst et al. (1990) between the
 median angular size ($\theta_{\rm med}$) and the radio flux 
$\theta_{\rm med}$ = 2$^{\prime\prime}$ S$^{0.30}_{\rm 1.4GHz}$ we estimated 
a median angular size between 0.95$^{\prime\prime}$ and 
1.54$^{\prime\prime}$ in the flux interval S$_{\rm 5GHz}$ = 0.05 - 0.25 mJy 
where we have $\sim$90\% of our sources (56/63).  
The 5 GHz flux density of our survey were transformed to 1.4 GHz 
using a radio spectral index $\alpha_{\rm r}$=0.4. 
Following Windhorst et al. (1990) we have assumed 
h($\theta$)=exp(-ln2 ($\theta$/$\theta_{\rm med}$)$^{0.62}$) for the 
integral angular distribution. 
We injected the sources simulated with the above recipes (each of them 
with known flux density and size) in the CLEANed sky images of the 
field. Then these sources were recovered from the image using 
the same procedure adopted to extract the real sources
(see Sect. 3.1) and binned in flux intervals. 
We repeated this simulation five times. 
From the comparison between the total number of sources detected in each 
bin (from the five simulations) and the total number of 
sources in the input sample in the same bin 
we calculated the correction factor $C$ to be applied to our observed source 
counts due to the incompleteness of the survey.  

\subsubsection{Source counts} 

For comparison with other 6 cm studies, 
the source counts are normalized to a non evolving Euclidean model which 
fits the brightest sources in the sky. At 6 cm the standard Euclidean 
integral counts are N($>$S$_{\rm 6~cm}$)=60$\times$S$_{\rm 6~cm}^{-1.5}$ sr$^{-1}$, 
with S$_{\rm 6~cm}$ expressed in Jy.  The results are given in Table 
~\ref{Table_counts} where, for each bin,  we report the flux interval, 
the average flux, 
the  number of sources detected (N), the correction factor $C$, 
the  number of sources after that the correction factor has been 
applied (N$_c$), 
the expected number  
of sources for a static Euclidean universe  (N$_{\rm exp}$) 
and the normalized  counts 
N$_{\rm c}$/N$_{\rm exp}$. The estimated error is N$_{\rm c}^{1/2}$/N$_{\rm exp}$. 
The average flux in each bin has been assumed equal to the geometric 
mean of the bin flux limits. Since in the flux interval sampled in 
our survey ($\sim$0.1 - 1.0 mJy) the best fit slope $\gamma$ of the 
normalized differential counts (dN/dS $\propto $S$^{-\gamma}$) 
is  $\gamma \sim$2.0 (Donnelly et al., 1987), this 
assumption is formally correct. In fact, only when   $\gamma$=2.0
the value of the average flux is equal to the geometric mean
(Windhorst et al., 1984). 

\begin{figure} 
\psfig{figure=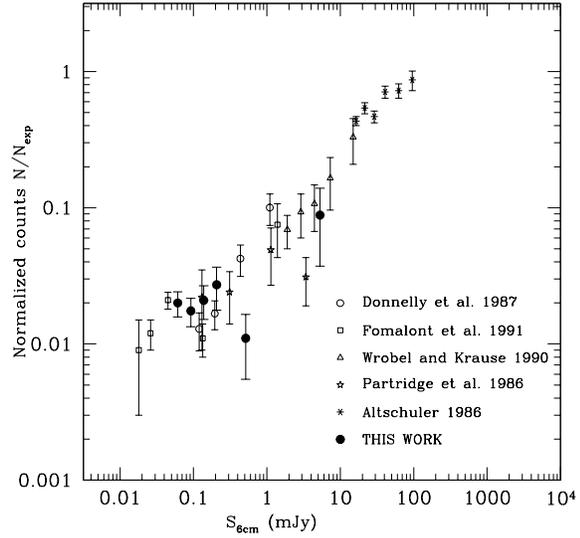,width=8cm} 
\caption[]{Differential 6 cm source counts normalized to a Euclidean 
Universe. Filled circles are our 6 cm counts in the Lockman Hole. The open 
points come from Donnelly et al. (1987);
 Fomalont et al. (1991); 
Wroble \& Krause (1990) and Partridge et al. (1986). 
The six points above 15 mJy come  from Altschuler (1986). } 

\label{F8}
\end{figure}

Our results are compared with previous work in Fig. ~\ref{F8}. 
In this figure we show only the source counts below $\sim$100 mJy. 
Above this flux the source counts are well known, with an initial 
steep rise between S$\simeq$10 and S$\simeq$1 Jy, a maximum excess with 
respect the Euclidean prediction between  S$\simeq$1.0 Jy and 
S$\simeq$0.1 Jy and a subsequent continuous convergence between 
 S$\simeq$100 and  S$\simeq$3 mJy. Below  S$\simeq$3 mJy there is a clear 
flattening in the slope of the 6cm radio source counts  
(see Windhorst et al. (1993) for a review of the radio source counts at 
different frequencies down to the microjansky level).

As illustrated in figure, there is a reasonably good agreement 
between the Lockman 
6 cm counts and  those obtained by other surveys in the same 
flux range, confirming that  there 
is no strong indication for a significant change in slope in the 6 cm counts 
between $\sim$0.1 and 1 mJy. Although the statistics is relatively poor, 
a possible flattening in the differential counts is instead suggested at 
fluxes below $\sim$ 0.1 mJy. 

\setcounter{table}{1}

  \begin{center}

   \begin{table*}
      \caption[]{Counts of radio sources}
         \label{Table_counts}
\begin{tabular}{crccccrc} 
&&&& \\ \hline
\multicolumn{3}{c}{Flux Density}    & Number  & $C$ & Corrected & N$_{\rm exp}$ & 
N$_{\rm c}$/N$_{\rm exp}$ \\  
\multicolumn{2}{c}{Range} & Average & Sources & & Number &  &          \\ 
\multicolumn{2}{c}{(mJy)}   & (mJy) &  (N) &   & Sources (N$_{\rm c}$) &  & 
($\times10^{-2}$)  \\

&&&& \\ \hline
  0.050  &   0.075  &  0.061 &  18 & 1.26 & 22.7 & 1135  & 2.00$\pm$0.42 \\ 
  0.075  &   0.113  &  0.092 &  17 & 1.10 & 18.7 & 1064  & 1.75$\pm$0.41 \\
  0.113  &   0.169  &  0.138 &  12 & 1.05 & 12.6 &  602  & 2.09$\pm$0.58 \\
  0.169  &   0.253  &  0.207 &  ~9 & 1.00 & ~9.0 &  330  & 2.72$\pm$0.91 \\
  0.253  &   1.281  &  0.569 &  ~4 & 1.00 & ~4.0 &  362  & 1.10$\pm$0.55 \\
  1.281  &  21.895  &  5.296 &  ~3 & 1.00 & ~3.0 &   34  & 8.82$\pm$5.09 \\ 
&&&& \\ \hline
\end{tabular}
\end{table*}

\end{center}

\section{Radio, optical and X-ray associations}

\subsection{Coincidences with 20 Centimeter Sources}
 
Forty-four of the 149  sources in the 4$\sigma$ radio sample 
at 20 cm published in de Ruiter et al. (1997) are within the 6 cm map. 
Two of these 
sources have more than one component (sources 71 and 99, see 
Table 1 in de Ruiter et al., 1997).   
When a compact 20 cm radio source is within 5$^{\prime\prime}$ 
from the 6 cm position, 
we accepted it as a real coincidence. Thus a 6 cm source was 
considered to be the counterpart of a 20 cm source only if the difference 
in position was significantly less than the one beam radius (the 20 cm 
map has been restored with a beam of $\sim$ 12$^{\prime\prime}$, 
see de Ruiter et al., 1997).
However, for extended 20 cm 
radio sources, we allowed a maximum distance of 10$^{\prime\prime}$
between the 20 cm and 6 cm positions. This is a necessary requirement in order 
not to miss 20-6 cm coincidences of low surface brightness sources. 
We find a total of 32 coincidences. 
Except for a few extended sources, the 20$-$6 cm positional correspondence is  
usually better than 3 arcsec. No other pair of 6-20 cm sources has been 
found at distances in the range 5-10 arcsec. 

Since the 6 and 20 cm surveys have two different beams (4$\times$4 arcsec 
and 12$\times$12 arcsec respectively), the spectral 
indices $\alpha$ (S$\propto \nu^{-\alpha}$) for all the  coincidences
were calculated using the total  20 cm fluxes reported in de Ruiter et 
al. (1997) and the integrated 6 cm fluxes obtained after convolving 
the 6 cm map with the same beam width as the 20 cm image. For 42 of the 
63 radio sources the differences between the  6 cm fluxes obtained 
from the maps with the two different beam sizes are smaller than 20\%, while 
only 4 sources show a difference between the two fluxes greater than a  
factor 1.7 (up to a factor 2.1). However these 4 sources have all a 
radio flux density lower than 0.1 mJy and have not been considered 
in the statistical analysis of the radio spectral index (see Sect. 5.1). 
The error in each 
calculated value of spectral index was computed by taking the 
quadrature sum of the relative errors in the two flux densities $S_1$ and 
$S_2$ : 

\begin{equation}
\sigma_{\alpha} = \sqrt{(\sigma_{S_1}/S_1)^2 + (\sigma_{S_2}/S_2)^2}/(ln~\nu_2 - ln~\nu_1)
\end{equation}
with $\nu_1$ = 1490 MHz and $\nu_2$ = 4860 MHz. 
For the 6 cm sources without a 20 cm counterpart, we calculated a 
4$\sigma$ 20 cm upper limit using $\sigma(\mu Jy)$ = 0.14r$^2$ - 1.78r +34.4,
where r is the distance (in arcmin) from the 20 cm image center 
(de Ruiter et al., 1997).  

The results of the 20$-$6 cm cross-correlation are summarized  in 
Table 3, where, for each 6 cm source we report the optical identification 
(Cols. 2--9, see below), the  name of the 20 cm 
counterpart (from Table 1 of 
de Ruiter et al., 1997), the 20 cm flux density (or 4$\sigma$ upper limit), 
the radio spectral index (or upper limit) and the distance between 
the two radio positions.  

The two  20 cm sources with more than one component 
(71 and 99) have been associated with  multi component 6 cm sources.  
In particular the triple source 71 has been associated
with the three component LOCK\_6cm J105148+573248. 
For this source we 
found a good agreement between the position of the components detected 
at the two radio frequencies. More complex is the 
situation for the other source (99).  This double source at 20 cm has been 
resolved into a four component source (LOCK\_6cm J105237+573104)
in the  6 cm map.  

\begin{figure} 
\hfill \psfig{figure=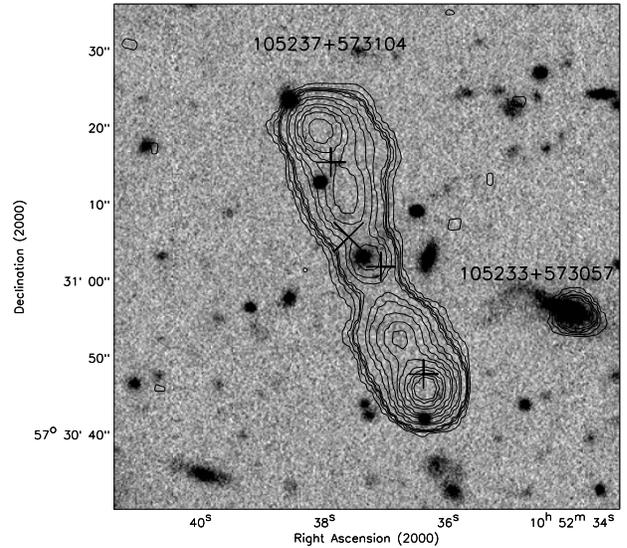,width=8cm} 
\caption[]{6 cm radio contour plot of the sources 
LOCK\_6cm J105237+573104 and LOCK\_6cm J105233+573057
superimposed on the V band CCD image. The three + symbols
represent the position of the 20 cm source 99 (99a, 99* and 
99b for increasing RA), while the symbol X represents the 
position of the X-ray source 116 identified by Lehmann et al. 
(2000) with the galaxy near the radio centroid.}
\label{F9}
\end{figure}

Moreover, a new, relatively bright radio source (LOCK\_6cm\_J105233+573057, 
S$_{\rm total}$=0.241 mJy) has been detected at a distance of 
25$^{\prime\prime}$  from the geometric center of the multiple source. 
This new radio source has a very unusual inverted radio spectral index 
($\alpha_{\rm r} < - 0.48$) and an optical counterpart that shows a clear 
extension in the direction of the multiple radio source. 
A 6 cm contour plot superimposed on the V band CCD image is shown in   
Fig. ~\ref{F9}.

For the two multiple sources we report in Table 3 only the spectral index 
derived from the total integrated flux. 

\subsection{Optical identification of 6 cm radio sources}

The entire 6 cm field is covered by V and I CCD data collected at the 
Canada-France-Hawaii Telescope (CFHT) with the UH8K camera 
(see Wilson et al. (2001) and Kaiser et al. (2001) 
for a description of these data). 
The approximate limits in the two bands 
are V$\sim$25.5 mag
and I$\sim$24.5 mag for point sources (Vega - magnitude).  
The CCD field is 30$^{\prime} \times 30^{\prime}$  
around the ROSAT ultra deep HRI and 6 cm field center.  
Moreover, a mosaic of 
K$^{\prime}$-band images (from the Omega-Prime camera on the 
Calar Alto 3.5-m telescope) covering the field in a non-uniform 
fashion with gaps in between is also 
available (see Lehmann et al. (2000) and Schmidt et al. (1998) for 
a summary of  the optical available data in the Lockman Hole). 
The limiting magnitudes of the K$^{\prime}$  
images is  K$^{\prime} \sim$ 20.2 mag.  The typical photometric error 
on the V, I and  K$^{\prime}$ magnitudes is $\sim$0.1 mag.  

\subsubsection{Identification in the  V and I bands}

For the optical identification of the 6 cm radio sources, we used the  
likelihood ratio technique described by Sutherland \& Saunders (1992). 
The mean off-set between the radio and optical positions estimated
in Sect. 3.2 has been subtracted from the radio positions to 
compute the positional offset  {\it r}. 
For a given optical candidate with magnitude {\it m} and 
positional offset {\it r} from the radio source position, we calculate
the probability {\it p} that the true source lies in an infinitesimal box
{\it r $\pm$ dr}/2, and in a magnitude interval {\it m $\pm$ dm}/2; on the
assumption that the positional offsets are independent of the
{\it optical} properties  {\it p} is given by:

\begin{equation}
p=q(m)~dm \times 2\pi~r~f(r)~dr,
\end{equation}

\ni where {\it f(r)} is the probability distribution function of the positional
errors, assumed to be equal in the two coordinates, with

\begin{equation}  
2\pi\int^{+\infty}_{0}f(r)~r~dr = 1
\end{equation}

\ni and {\it q(m)} is the expected distribution as a function of magnitude 
of the optical counterparts. 

The likelihood ratio {\it LR} is defined as the ratio between the probability
that the source is the correct identification
 and the corresponding probability for a 
background, unrelated object:

\begin{equation}
LR=\frac{q(m) f(r)}{n(m)}
\end{equation}

\ni where {\it n(m)} is the surface density 
of background objects with magnitude {\it m}.

The presence or absence of other optical candidates for the same radio source
provides {\it additional information} to that contained in {\it LR}.
Thus a self-consistent formula has been developed for the reliability
of each individual object, taking this information into account
(Sutherland \& Saunders, 1992).
The reliability $Rel_j$ for object {\it j} being the correct identification
is:

\begin{equation}
Rel_j=\frac{(LR)_j}{\Sigma_i (LR)_i + (1-Q)}
\end{equation}

\ni where the sum is over the set of all candidates 
for this particular source 
and Q is the probability that the optical counterpart of the source is
brighter than the magnitude limit of the optical catalogue 
($Q = \int^{m_{lim}} q(m)~dm $).

In order to derive an estimate for {\it q(m)} 
we have first counted all objects in the 
optical catalogue within a fixed radius around each source ({\it total(m)}). 
This distribution has then been background subtracted 

\begin{equation}
real(m) = [total(m) - n(m)*N_{radio~sources}*\pi*radius^2]
\end{equation}

\ni and normalized
to construct the distribution function of `real' identifications :  

\begin{equation}
q(m) = \frac{real(m)}{\Sigma_i real(m)_i} \times Q
\end{equation}

\ni where $\Sigma_i$ is the sum over all the magnitude 
bins of the distribution, $i.e.$ 
is the total number of objects in the $real(m)$ distribution. 

\begin{figure} 
\hfill \psfig{figure=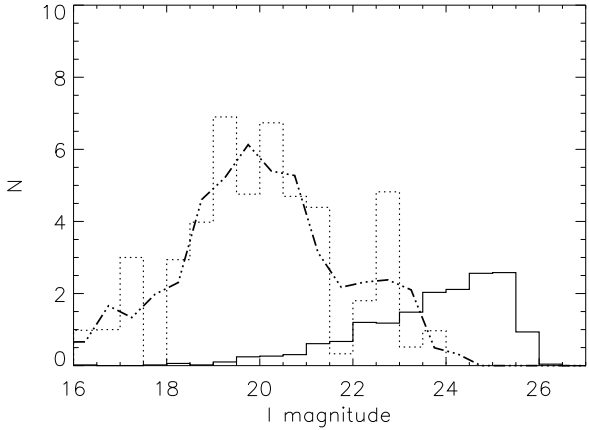,width=9cm} 
\caption[]{Magnitude distributions of 
background objects (solid line) and 'real' detections 
({\it real(m)} ,dotted line) estimated from 
the optical objects detected in the I band within a radius of 2 
arcsec around each radio sources.  The smooth curve fitted to the  
{\it real(m)} distribution (dot dot dot dashed line) has been 
normalized according to Eq. (9) and then used as 
input in the 
likelihood calculation.}

\label{F10}
\end{figure}

In order to maximize the statistical significance of the over density due 
to the presence of the optical counterparts, we have adopted for the 
radius the minimum size which assures that most of the possible 
counterparts are included within such radius. On the basis of the positional 
uncertainties shown in Fig. \ref{F6}, we have adopted a radius 
of 2 arcsec. Fig. \ref{F10} shows the resulting  {\it real(m)} distribution
(dotted line) together with the expected distribution of background objects
(solid line), 
$i.e$ objects 
unrelated to the radio sources . The smooth curve fitted to the  
{\it real(m)} distribution (dot dot dot dashed line) has been 
normalized according to Eq. (9) and then used as 
input in the 
likelihood calculation. The total number of objects in the  
{\it real(m)} distribution is 48.8 $\pm$ 8.7. Since the number of 
radio sources is 63, this corresponds to an expected fraction of 
identifications 
above the magnitude limit of the optical catalogue of the order of 
(77 $\pm$ 14) \%. On this basis we adopted   $Q$ = 0.8. 
This value is also in agreement  
with the results obtained by Hammer et al. (1995) for the 
radio sources in the Canada-France Redshift Survey and by 
Fomalont et al. (1997) and Richards et al. (1998) for the radio sources 
in the Hubble Deep Field (HDF). 
Using I band CCD data with 
a magnitude limit of I $\sim$ 24.5 mag Hammer et al. found 32 optical 
identifications for 36 radio sources with S$_{\rm 5.0~GHz}>16\mu$Jy, 
while Richards et al. found 26 optical counterparts down to I=25 mag
for 29 radio sources with  S$_{\rm 8.5~GHz}>9\mu$Jy. 

However, to check how 
this assumption could affect our results, we repeated the likelihood ratio 
analysis using different values of $Q$ in the range 0.5--1.0. No 
substantial difference in the final number of identifications and in the 
associated reliability (see below) has been found. 

As probability distribution of positional errors we adopted a 
Gaussian distribution with 
standard deviation, $\sigma$, which takes into account the combined effect 
of the radio and the optical positional uncertainties:

\begin{equation}  
f(r)=\frac{1}{2 \pi \sigma^2} exp(\frac{-r^2}{2 \sigma^2})
\end{equation}

\ni For each source the value of $\sigma$  used is  
the average value between 
$\sigma_x = \sqrt{er_{op}^2+\sigma_{\alpha}^2}$ 
and $\sigma_y = \sqrt{er_{op}^2+\sigma_{\delta}^2}$, where 
$er_{op}$ is the error on the optical position (we assumed a  
value of 0.5 arcsec), while $\sigma_{\alpha}$ and  $\sigma_{\delta}$ are 
the radio 
positional errors in RA and DEC reported in Table 1.

Having determined the values of  {\it q(m)}, {\it f(r)} and {\it n(m)},  we 
computed the $LR$ value for all the optical sources within a distance 
of 5 arcsec from the radio position. Once that the $LR$ values have 
been computed 
for all the optical candidates, one has to choose the best threshold 
value for $LR$ ($L_{\rm th}$) to discriminate between spurious and real 
identifications. The choice of $L_{\rm th}$ depends on two factors: first, it 
should be small enough to avoid missing many real 
identifications and having a rather incomplete sample. 
Secondly, $L_{\rm th}$ should 
be large enough to keep the number of spurious identifications  
as low as possible and to increase the reliability.  

As  $LR$  threshold we adopted $L_{\rm th}$=0.2. 
With this value, according to Eq. (7) and considering that 
our estimate for Q is 0.8, all the optical counterparts of radio sources 
with only one identification (the majority in our sample)
and $LR>LR_{\rm th}$ have a reliability greater than 0.5. This choice also 
approximately maximizes the sum of sample reliability and completeness.

With this threshold value we find 56 radio sources with a likely 
identification (three of which have two optical candidates with 
 $L_{\rm th}>$0.20 for a total of 59 optical candidates with 
 $L_{\rm th}>$0.20).  

The reliability (Rel) of each of 
these optical identifications (see Eq. 7) is always 
high ($>$95 per cent for most 
of the sources), except for the few cases 
where more than one optical candidate 
with  $L_{\rm th}>$0.20 is present for the same radio source. 

The number of expected real identifications (obtained adding  the 
reliability of all the objects with   $L_{\rm th}>$0.20) is about 53
$i.e$ we expect that about 3 of the 56 proposed radio-optical 
associations may be  
spurious positional 
coincidences. Similar results are obtained using the catalogue in the 
V band. The results of the optical identification are summarized 
in Table 3. For each radio source we report all the optical counterparts 
within 3 arcsec and with a magnitude I$\leq$24.5  plus two objects which, 
having I$>$24.5 mag, are too faint for a reliable determination of their 
likelihood ratio, but we however considered as likely identification 
because of their small distance (lower than 0.6 arcsec) from the 
radio position  (LOCK\_6cm\_J105158+573330 and LOCK\_6cm\_105259+573226). 

In the three cases in which more than one optical object with 
$LR>LR_{\rm th}$ have been found associated to the same radio source, 
we assumed the object with the 
highest Likelihood Ratio value as the counterpart of this radio source. 

\begin{sidewaystable*}

\footnotesize

\caption[]{Optical, near infrared and 20 cm  counterparts of the 6 cm sources in the Lockman Hole}

\label{catalog}

\begin{tabular}{lrrrcrrrclrcrcc} 

&&&& \\ \hline
NAME  &  \multicolumn{1}{c}{V$^a$} &    \multicolumn{1}{c}{I$^a$}  &  \multicolumn{1}{c}{K$^{\prime~a}$} &  $\Delta$ & $\Delta$RA & $\Delta$DEC & \multicolumn{1}{c}{LR} & Rel. &  20cm &  F20 & F20$_{\rm err}$ & \multicolumn{1}{c}{$\alpha_{\rm r}$} &  $\alpha_{\rm r~err}  $ & $\Delta6-20$    \\
      &    &              &                     & ($^{\prime\prime}$) & ($^{\prime\prime}$) & ($^{\prime\prime}$) &   &  & Nr. & mJy & & & &  ($^{\prime\prime}$) \\ \hline

 LOCK\_6cm\_J105135+572739  &  22.8  & 21.1   & 18.2 &  1.02 &  $-$0.97  &  0.30   &   8.66  &  0.98 &  59   &    0.190 &   0.030 &    0.63 &   0.21   &  2.6  \\
 LOCK\_6cm\_J105136+572959  &0.0$^b$ & 20.4   & 17.8 &  0.81 &  $-$0.81  &  0.09   &  23.83  &  0.98 &  61   &    0.399 &   0.050 &    1.24 &   0.19   &  3.9  \\
 LOCK\_6cm\_J105136+573302  &0.0$^b$ & 21.5   & 18.8 &  1.15 &  $-$0.67  &  0.94   &   2.69  &  0.93 &  60   &    0.447 &   0.040 &    1.01 &   0.15   &  0.5  \\
                            &0.0$^b$ & 21.5   &$>$20.2 &  2.70 &    0.27   &  2.69   &   0.01  &  0.00 &  60   &    0.447&  0.040 &    1.01 &   0.15   &  0.5  \\
 LOCK\_6cm\_J105137+572940  &0.0$^b$ & 20.7   & 17.4 &  1.42 &  $-$1.41  &  0.13   &   5.72  &  0.97 &  63   &    2.210 &   0.080 &    0.49 &   0.03   &  0.4  \\
 LOCK\_6cm\_J105143+572938  &0.0$^b$ & 15.4   & 13.5 &  1.62 &  $-$0.60  &  2.55   &  24.66  &  0.99 &  69   &    0.421 &   0.040 &    0.52 &   0.19   &  3.3  \\
 LOCK\_6cm\_J105143+573213  &0.0$^b$ &$>$24.5 &      &       &           &         &      -  &     - &       & $<$0.126 &         & $<$0.65 &          &         \\
 LOCK\_6cm\_J105148+573248T &0.0$^b$ & 21.4   & 18.0 &  0.72 &  $-$0.40  &  0.59   &   7.71  &  0.97 &  71*  &   15.390 &   0.580 &    0.95 &   0.03   &  0.2  \\
 LOCK\_6cm\_J105150+573245  &0.0$^b$ & 18.0   &      &  1.41 &  $-$0.50  &  1.31   &   1.87  &  0.90 &       &  $<$0.126 &         &$<-$0.06 &          &	        \\
 LOCK\_6cm\_J105150+572635  &  25.4  & 22.8   & 19.3  &  1.03 &  $-$0.87  & $-$0.55 &  33.79  &  0.99 &  72   &    0.216 &   0.060 &   0.00 &    0.26  &  2.0  \\
 LOCK\_6cm\_J105156+573322  &  22.1  & 20.7   &$>$20.2&  1.54 &   0.37    & $-$1.50 &   4.67  &  0.96 &       & $<$0.131 &         & $<$0.70 & 	       &	        \\
 LOCK\_6cm\_J105158+572330  &$>$25.5 & 25.3   &       &  0.59 &  $-$0.57  &  0.14   &   0.00  &  0.00 &  74   &    0.195 &   0.030 &    0.47 &   0.17  &  0.6  \\
 LOCK\_6cm\_J105210+573317  &  23.9  & 22.7   & 19.4  &  1.82 &  $-$0.13  & $-$1.81 &   0.41  &  0.50 &       & $<$0.131 &         & $<$0.69 &         &	        \\
                            &  23.5  & 22.3   & 19.6  &  2.02 &  $-$1.34  &    1.52 &   0.22  &  0.26 &       & $<$0.131 &         & $<$0.69 &         &	        \\   
 LOCK\_6cm\_J105211+572907  &  25.5  & 23.6   & 19.3  &  0.67 &  $-$0.64  & $-$0.20 &   0.77  &  0.79 &  81   &    1.455 &   0.060 &    0.97 &   0.04  &  0.3  \\
 LOCK\_6cm\_J105212+572453  &  21.6  & 19.8   & 17.1  &  0.52 &  $-$0.50  & $-$0.14 &  70.21  &  0.99 &  83   &    0.268 &   0.030 &    0.97 &   0.14  &  3.3  \\
 LOCK\_6cm\_J105213+572650  &  23.7  & 20.4   & 16.8  &  0.07 &  $-$0.07  &  0.02   &  57.99  &  0.99 &       & $<$0.115 &         & $<$0.79 &         &         \\
 LOCK\_6cm\_J105216+573529  &  18.7  & 17.5   & 15.1  &  0.20 &  $-$0.20  &  0.00   & 287.12  &  0.99 &  85   &    0.206 &   0.040 &    1.39 &   0.32  &  4.0  \\
 LOCK\_6cm\_J105225+573322  &  21.3  & 18.6   & 15.7  &  0.64 &  $-$0.64  & $-$0.02 & 231.92  &  0.99 &  89   &    4.515 &   0.160 &    0.81 &   0.03  &  0.5 \\
 LOCK\_6cm\_J105226+573711  &  24.8  & 23.4   &       &  2.18 &  1.97     & $-$0.92 &   0.02  &  0.09 &       & $<$0.167 &         & $<$1.21 &         &         \\
 LOCK\_6cm\_J105227+573832  &  24.2  & 22.6   &       &  1.79 &  $-$1.74  &  0.42   &   0.31  &  0.60 &       & $<$0.182 &         & $<$1.00 &         &	        \\
 LOCK\_6cm\_J105229+573334  &  25.1  & 24.1   &$>$20.2&  1.68 &  $-$1.64  &  0.33   &   0.04  &  0.15 &       & $<$0.137 &         & $<$0.98 &         &	        \\
 LOCK\_6cm\_J105231+573027  &  22.1  & 19.1   & 16.0  &  1.00 &  $-$0.60  &  0.80   &  64.56  &  0.99 &       & $<$0.121 &         & $<$0.62 &	       &	        \\
 LOCK\_6cm\_J105231+573204  &  22.2  & 20.1   & 17.2  &  2.28 &  $-$0.03  &  2.28   &   0.68  &  0.48 &       & $<$0.128 &         & $<$1.15 & 	       &	        \\
                            &  23.6  & 22.2   & 20.1  &  1.64 &     1.64  & $-$0.05 &   0.53  &  0.37 &       & $<$0.128 &         & $<$1.15 & 	       &	        \\
 LOCK\_6cm\_J105231+573501  &  21.4  & 19.0   & 17.8  &  0.76 &  $-$0.44  & $-$0.63 &  94.10  &  0.99 &  95   &    0.203 &   0.040 &    0.59 &   0.19  &  3.4  \\
                            &  21.2  & 18.9   & 17.5  &  2.68 &     2.68  & $-$0.03 &   0.26  &  0.00 &  95   &    0.203 &   0.040 &    0.59 &   0.19  &  3.4  \\
 LOCK\_6cm\_J105233+573057  &  20.2  & 18.5   & 16.0  &  0.89 &  $-$0.87  & $-$0.17 & 144.83  &  0.99 &       & $<$0.124 &         &$<$$-$0.48 &       &	        \\
 LOCK\_6cm\_J105234+572643  &  21.8  & 20.3   & 18.1  &  0.43 &  $-$0.34  &  0.27   &  47.11  &  0.99 &       & $<$0.116 &         & $<$0.75 &	       &	        \\
 LOCK\_6cm\_J105235+572640  &  21.3  & 19.7   & 17.0  &  1.85 &   1.65    & $-$0.84 &   7.96  &  0.98 &       & $<$0.116 &         & $<$0.75 &	       &	        \\
 LOCK\_6cm\_J105237+572147  &  21.8  & 20.0   & 17.2  &  1.26 &  $-$0.61  & $-$1.11 &  17.86  &  0.99 &       & $<$0.116 &         & $<$0.63 &	       &	        \\
 LOCK\_6cm\_J105237+573104T &  21.8  & 19.0   & 16.0  &  0.23 &   0.10    &  0.20   & 203.41  &  0.99 &  99*  &   59.452 &   1.910 &    1.00 &   0.03  &  4.1  \\  
 LOCK\_6cm\_J105238+572221  &$>$25.5 &$>$24.5 &       &       &           &         &         &       &       & $<$0.115 &         & $<$0.76 & 	       &	        \\
 LOCK\_6cm\_J105238+573321  &  22.4  & 20.3   & 17.5  &  1.78 &  $-$1.48  & $-$1.00 &   2.38  &  0.92 &       & $<$0.138 &         & $<$0.97 &         &         \\
 LOCK\_6cm\_J105239+572430  &  17.8  & 17.2   & 16.0  &  1.12 &  $-$0.84  & $-$0.73 & 237.62  &  0.99 &  100  &    0.144 &   0.030 &    0.92 &   0.22  &  1.3  \\
 LOCK\_6cm\_J105241+572320  &  23.3  & 20.9   & 17.5  &  0.13 &   0.00    & $-$0.13 &  29.55  &  0.99 &  101  &    1.724 &   0.070 &    0.59 &   0.04  &  0.8  \\

\hline

\end{tabular}       

$^a$ Magnitudes in all bands are Vega magnitudes 

\ni $^b$ For this source the V magnitude is not available due to a problem with the electronic readout in the CCD \\

\end{sidewaystable*}

\setcounter{table}{2}

\begin{sidewaystable*}

  \caption{\it continued}
  \label{catalog}
\begin{tabular}{lrrrcrrrclrcrcc} 
&&&& \\ \hline
NAME  & \multicolumn{1}{c}{V$^a$} &  \multicolumn{1}{c}{I$^a$}   & \multicolumn{1}{c}{K$^{\prime~a}$} &  $\Delta$ & $\Delta$RA & $\Delta$DEC & \multicolumn{1}{c}{LR} & Rel. &  20cm &  F20 & F20$_{\rm err}$ & \multicolumn{1}{c}{$\alpha_{\rm r}$} &  $\alpha_{\rm r~err}  $ & $\Delta6-20$    \\
      &    &       &                           & ($^{\prime\prime}$) & ($^{\prime\prime}$) & ($^{\prime\prime}$) &   &  & Nr. & mJy & & & &  ($^{\prime\prime}$) \\ \hline

 LOCK\_6cm\_J105242+571915  &  18.0 & 16.7   & 14.7 &  1.10 &  $-$0.20  &  1.08   & 297.78  &  0.99 &  103  &    0.246 &   0.030 &    0.35 &   0.17   &  1.2  \\
 LOCK\_6cm\_J105245+573615  &  21.8 & 19.2   &      &  0.61 &   0.33    & $-$0.52 & 135.25  &  0.99 &  104  &    0.458 &   0.060 &    0.46 &   0.12   &  2.8  \\
 LOCK\_6cm\_J105249+573626  &  25.8 & 22.0   & 18.9 &  0.27 &   0.27    & $-$0.02 &   6.94  &  0.97 &       & $<$0.170 &         & $<$1.00 &	      &	        \\
 LOCK\_6cm\_J105252+572859  &  17.8 & 16.5   & 14.2 &  0.17 &  $-$0.17  &  0.03   & 797.05  &  0.99 &  105  &    0.202 &   0.030 &    0.61 &   0.17   &  3.4  \\
 LOCK\_6cm\_J105254+572341  &  23.1 & 20.8   & 18.2 &  0.67 &  $-$0.34  & $-$0.58 &  16.99  &  0.99 &       & $<$0.115 &         &$<$$-$0.06 &        &         \\
 LOCK\_6cm\_J105255+571950  &  24.4 & 22.6   & 18.1 &  0.53 &  $-$0.51  & $-$0.14 &   6.65  &  0.97 &  109  &    2.692 &   0.090 &    0.93 &   0.03   &  0.9  \\
 LOCK\_6cm\_J105259+573226  &$>$25.5& 24.6   & 20.1 &  0.33 &   0.30    &  0.13   &   0.00  &  0.00 &  113  &    0.285 &   0.040 &    0.38 &   0.13   &  2.4  \\
 LOCK\_6cm\_J105301+572521  &  22.9 & 20.0   & 17.2 &  0.08 &   0.00    &  0.08   &  92.84  &  0.99 &  114  &    0.151 &   0.030 &    0.38 &   0.19   &  0.5  \\
 LOCK\_6cm\_J105301+573333  &  21.8 & 19.2   & 16.3 &  1.22 &   1.21    & $-$0.14 &  43.30  &  0.99 &       & $<$0.152 &         & $<$0.97 &	      &	        \\
 LOCK\_6cm\_J105303+573429  &  25.8 & 23.3   & 18.5 &  0.92 &   0.80    & $-$0.45 &   0.47  &  0.70 &       & $<$0.161 &         & $<$0.74 &	      &	        \\
 LOCK\_6cm\_J105303+573526  &  21.8 & 19.8   & 17.1 &  0.93 &  $-$0.10  & $-$0.92 &  34.49  &  0.99 &       & $<$0.160 &         & $<$0.71 &	      &	        \\
 LOCK\_6cm\_J105303+573532  &  22.0 & 20.3   & 17.1 &  0.70 &   0.64    &  0.30   &  30.62  &  0.99 &  116  &    0.366 &   0.040 &    0.81 &   0.18   &  1.0  \\
 LOCK\_6cm\_J105304+573055  &  22.2 & 19.0   & 15.8 &  0.44 &  $-$0.44  & $-$0.03 & 167.26  &  0.99 &  119  &    0.440 &   0.040 &    0.69 &   0.10   &  0.3  \\
 LOCK\_6cm\_J105308+572223  &  20.8 & 18.7   &      &  0.10 &  $-$0.10  &  0.00   & 364.69  &  0.99 &  121  &    0.340 &   0.030 &    0.61 &   0.10   &  0.3  \\ 
 LOCK\_6cm\_J105308+573436  &  20.8 & 18.9   & 16.5 &  0.48 &  $-$0.20  & $-$0.44 & 151.27  &  0.99 &       & $<$0.166 &         & $<$0.73 &	      &	        \\
 LOCK\_6cm\_J105310+573434  &  22.9 & 21.9   & 19.2 &  1.89 &  $-$0.94  & $-$1.64 &   0.27  &  0.47 &       & $<$0.167 &         & $<$0.75 &	      &	        \\
                            &  23.2 & 21.7   & 19.6 &  2.16 &     1.71  &    1.33 &   0.10  &  0.18 &       & $<$0.167 &         & $<$0.75 &	      &	        \\
 LOCK\_6cm\_J105312+573111  &  24.6 & 21.3   & 17.9 &  0.59 &  $-$0.57  & $-$0.16 &   7.95  &  0.98 &  126  &    0.232 &   0.040 &    0.74 &   0.16   &  1.4  \\   
 LOCK\_6cm\_J105313+572507  &  25.9 & 23.7   & 17.5 &  0.66 &  $-$0.17  & $-$0.64 &   0.70  &  0.77 &       & $<$0.124 &         & $<$0.41 &	      &	        \\
                            &  24.7 & 23.0   & 18.0 &  2.48 &  1.28     & $-$2.13 &   0.01  &  0.01 &       & $<$0.124 &         & $<$0.41 &	      &	        \\
 LOCK\_6cm\_J105314+572448  &  23.3 & 20.4   &      &  0.13 &   0.10    & $-$0.08 &  55.04  &  0.99 &       &    0.124 &         &$<-$0.14 &	      &	        \\
 LOCK\_6cm\_J105314+573020  &  24.3 & 22.8   & 18.8 &  0.99 &  $-$0.94  & $-$0.30 &   2.08  &  0.88 &  127  &    0.292 &   0.040 &    0.62 &   0.15   &  2.4  \\ 
                            &  24.4 & 23.3   & 19.8 &  1.59 &  $-$1.31  &    0.91 &   0.09  &  0.04 &  127  &    0.292 &   0.040 &    0.62 &   0.15   &  2.4  \\      
 LOCK\_6cm\_J105316+573551  &  19.1 & 18.2   & 16.7 &  0.86 &  $-$0.40  &  0.77   & 129.49  &  0.99 &  128  &    0.255 &   0.050 &    0.64 &   0.20   &  1.6  \\     
 LOCK\_6cm\_J105317+572722  &  24.2 & 22.9   &$>$20.2&  0.95 &  $-$0.91  & $-$0.27 &  2.19  &  0.90 &       & $<$0.132 &         & $<$1.10 &	      &	        \\
 LOCK\_6cm\_J105322+573652  &  21.8 & 19.2   & 16.4 &  0.53 &  $-$0.50  & $-$0.16 & 128.50  &  0.99 &       & $<$0.202 &         & $<$1.00 &	      &	        \\
 LOCK\_6cm\_J105325+572911  &  17.9 & 17.2   & 15.7 &  1.21 &  $-$0.87  &  0.84   & 169.03  &  0.99 &  134  &    0.552 &   0.040 &    0.51 &   0.13   &  1.3  \\
 LOCK\_6cm\_J105327+573316  &  22.2 & 20.5   & 18.1 &  0.61 &  $-$0.54  & $-$0.30 &  40.77  &  0.99 &       & $<$0.173 &         & $<$0.42 &	      &	        \\
 LOCK\_6cm\_J105328+573535  &  22.2 & 19.4   & 16.4 &  0.55 &  $-$0.44  &  0.33   &  87.01  &  0.99 &       & $<$0.196 &         & $<$0.82 &	      &	        \\
 LOCK\_6cm\_J105335+572157  &  25.0 & 24.4   &      &  2.06 &     0.24  &  2.05   &   0.00  &  0.00 &       & $<$0.139 &         & $-$0.01 &          &	        \\
 LOCK\_6cm\_J105335+572921  &  23.3 & 20.0   & 17.0 &  0.46 &  $-$0.34  &  0.31   &  71.86  &  0.99 &  138  &    0.214 &   0.040 &    0.53 &   0.18   &  1.4  \\
 LOCK\_6cm\_J105342+573026  &$>$25.5& 23.9   & 20.1 &  0.95 &     0.27  &  0.91   &   0.24  &  0.55 &  142  &    0.274 &   0.040 &    1.08 &   0.20   &  1.2  \\
 LOCK\_6cm\_J105347+573349  &  22.6 & 22.3   &      &  0.74 &  $-$0.70  &  0.22   &   4.64  &  0.96 &       & $<$0.203 &         & $<$0.21 &  	      &	        \\
 LOCK\_6cm\_J105348+573033  &  23.3 & 21.2   & 18.5 &  0.71 &  $-$0.64  &  0.31   &  14.96  &  0.99 &       & $<$0.180 &         & $<$0.80 &  	      &	        \\

\hline                                                                          
\end{tabular}       

$^a$ Magnitudes in all bands are Vega magnitudes

\end{sidewaystable*}

\normalsize

\subsubsection{Identification in the  K$^{\prime}$ band}

As said above, the  K$^{\prime}$ band data cover the 6 cm field in a  
non-uniform fashion. For 12 of the 63 radio sources  K$^{\prime}$ band data
are not available. Since all the 51 radio sources with  available
K$^{\prime}$ band data do have an optical identification in the I band, 
we looked for K$^{\prime}$ counterparts using a maximum distance 
of 1.0 arcsec from the optical position. 
We found a  K$^{\prime}$ counterpart for 49 of the 51 
radio sources. The same results is obtained using a search radius of 
2 arcsec. 

A summary of  the results of our identifications 
is given in Table 3. For each radio source, we give the V, I  and 
 K$^{\prime}$ magnitude (when available), the total distance 
and the distance in RA and DEC 
between the radio and the optical position  and  
the Likelihood Ratio and reliability values obtained using the I band 
catalogue. For the radio sources with more than one optical counterpart
we assumed as the real identification  the optical source with the 
highest reliability.  A blank field in the   K$^{\prime}$ magnitude 
column means no data available. 

In summary, considering as  good identifications also the two 
sources fainter than I$\sim$24.5 mag, we have a proposed  identification 
for 58 of the 63 radio sources (92 per cent): $\sim$54 of these 
are likely to be the correct identification. Moreover, for a subsample of 
51 radio sources (all of them with an optical identification in I) we have 
information also in the   K$^{\prime}$ band. For this subsample we 
found  K$^{\prime}$ counterparts for 49 sources (96 per cent of 
identifications).

\subsection{X-ray counterparts}

The area from which we have extracted our radio catalogue (10 arcmin radius) 
has been covered by the ROSAT Ultra Deep Survey (about 1 million seconds
of exposure with the HRI) reaching a limiting flux of about 
1.2$\times10^{-15}$ erg cm$^{-2}$ s$^{-1}$
in the 0.5 - 2.0 keV energy band. The HRI catalogue (Lehmann et al., 2001)
lists 54 sources in this area. A cross- correlation 
was performed between the position of the 54 X-ray sources 
and the 63 radio sources detected at 6 cm. 
We find 8 reliable radio/X-ray associations with 
a positional difference smaller than 5 arcsec.
An additional radio/X-ray association is found with
the PSPC source 116 (Hasinger et al., 1998) which was not detected 
with the HRI. At these fluxes (S$_{\rm 0.5-2.0~keV}\sim$1.2$\times10^{-15}$ 
erg cm$^{-2}$ s$^{-1}$, S$_{\rm 6~cm}\sim50\mu$Jy) 
the surface densities of radio and X-ray
sources are similar and the coincidences between the samples selected at
radio and X-ray frequencies are about 15\%. All these nine 
radio/X-ray sources
have been optically identified by 
Lehmann et al. (2000) and Lehmann et al. (2001).
The radio/X-ray associations are given in Table~\ref{TableRX}. 
For each radio source we 
report the radio flux, the radio spectral index $\alpha_{\rm r}$, 
the I and K$^{\prime}$ band magnitude, the name of the X-ray counterpart, 
the distance 
between the radio and X-ray position, the ROSAT HRI (except for source 116 
for which we have the ROSAT-PSPC flux) 0.5-2.0 keV flux of 
the X-ray source in units of  10$^{-14}$ erg cm$^{-2}$ s$^{-1}$, the optical 
classification (1 for Type I AGN, 2 for Type II AGN and 3 for cluster),  
the redshift and the radio luminosity in W/Hz  (H$_0$=70 km s$^{-1}$ 
Mpc$^{-1}$, $\Omega_M$=0.3,  $\Omega_{\lambda}$=0.7).

\setcounter{table}{3}
   \begin{table*}
      \caption[]{Radio/X-ray associations}
         \label{TableRX}
\begin{tabular}{lcrcccccclc} 
&&&& \\ \hline
Radio source & F$_{\rm 6~cm}$ &$\alpha_{\rm r}$ & I & K$^{\prime}$  & Nr. X & $\Delta$R-X & Fx$^a$ & Type & \multicolumn{1}{c}{z} & Log L$_{\rm 5~GHz}$\\
             & (mJy)     &           &   &               &       & ($^{\prime\prime}$) &  & &  & (W/Hz)  \\ \hline
LOCK\_6cm\_J105148+573248T &~5.03 &0.95      & 21.4 & 18.0  & ~12   & 0.8 & 0.72 & 2 & 0.990 & 25.41\\
LOCK\_6cm\_J105237+573104T &16.45 &1.00      & 19.0 & 16.0  & 116   & 3.5 & 0.57 & 2 & 0.708$^b$ & 25.56\\ 
LOCK\_6cm\_J105239+572430  &~0.06 &0.59      & 17.2 & 16.0  & ~32   & 0.8 & 7.02 & 1 & 1.113 & 23.61\\ 
LOCK\_6cm\_J105252+572859  &~0.10 &0.41      & 16.5 & 14.2  & 901   & 1.2 & 0.12 & 2 & 0.204 & 22.07\\
LOCK\_6cm\_J105254+572341  &~0.13 &$<-$0.06  & 20.8 & 18.2  & 513   & 1.5 & 0.43 & 1 & 0.761 & 23.54\\
LOCK\_6cm\_J105303+573532  &~0.14 &0.81      & 20.3 & 17.1  & 827   & 2.8 & 0.18 & 2 & 0.249 & 22.42\\ 
LOCK\_6cm\_J105316+573551  &~0.09 &0.64      & 18.2 & 16.7  & ~~6   & 0.7 & 9.32 & 1 & 1.204 & 23.87\\ 
LOCK\_6cm\_J105328+573535  &~0.08 &$<$0.82   & 19.4 & 16.4  & 815   & 4.3 & 0.18 & 3 & 0.700$^b$ & 23.24\\
LOCK\_6cm\_J105348+573033  &~0.09 &$<$0.80   & 21.2 & 18.5  & 117   & 1.3 & 0.66 & 2 & 0.780 & 23.41\\ 
& & &  & \\
\multicolumn{11}{l}{\em New radio/X-ray associations found with the XMM-Newton source list} \\ 
& & &  & \\
LOCK\_6cm\_J105238+573321  &~0.05 & $<$0.97  & 20.3  & 17.5 & 52061 & 2.3 & 0.17 & 2 & 0.707$^b$ & 23.04 \\
LOCK\_6cm\_J105255+571950  &~1.02 & 0.93     & 22.6  & 18.1 & 52149 & 1.8 & 0.07 & 2 & 1.450 & 25.13 \\
LOCK\_6cm\_J105304+573055  &~0.18 & 0.69     & 19.0  & 15.8 & 52198 & 1.6 & 0.05 & 2 & 0.805 & 23.74 \\
LOCK\_6cm\_J105347+573349  &~0.17 & $<$0.21  & 22.3  &      & 52048 & 2.5 & 0.21 & 1 & 2.586 & 24.96 \\   
\hline
\end{tabular}

\footnotesize
$^a$ Unit of 10$^{-14}$ erg cm$^{-2}$ s$^{-1}$. 
Note: The ROSAT/X-ray data, including spectroscopic classification and 
redshift  are all taken from Lehmann et al. (2001), except for the X-ray 
source 116 for which the references are Hasinger et al. (1998) and 
Lehmann et al. (2000).  
The XMM/X-ray data are from Lehmann et al. (2003), in preparation.  \\
\ni $^b$ The sources J105237+573104T and J105238+573321, although with the same redshift, 
are not members of the cluster at z=0.700 associated with the radio source J105328+573535. However, these three radio/X-ray sources suggest the presence 
of an overdensity extending over at least a few Mpc at z$\sim$0.7. 

\normalsize
\end{table*}

Recently the Lockman Hole has been observed also with the XMM-Newton 
satellite (Hasinger et al., 2001; Lehmann et al., 2002). 
The center of the XMM-Newton observation 
is coincident with the ROSAT HRI field center and with the 6 cm 
survey discussed in this paper. 
A total of $\sim$100 ksec good exposure time has been accumulated, 
reaching a flux limit of 3.8$\times$ 10$^{-16}$ erg cm$^{-2}$ s$^{-1}$ 
in the 0.5-2.0 keV band.  
Within an off-axis angle of 10 arcmin ($i.e.$  the same region of the 
sky that we used to extract the 6 cm radio sources) a total of 106 
sources have been detected (see Lehmann et al., 2002). In addition to the 
9 radio/X-ray associations already found with the ROSAT data, 
there are 
four new radio/X-ray associations with the XMM-Newton source list.
The percentage of XMM/X-ray sources with a radio counterpart is therefore 
$\sim$12.3 per cent (13/106). 
The properties of the new radio/XMM associations 
are reported in the last 4 lines of Table~\ref{TableRX}.

It is interesting to note that out 
of 13 radio/X-ray  associations, eight (62\%) are 
classified as Type II AGN.  
Eight of the 13 radio sources are detected at both 21 and 6 cm and all
of them have steep spectral indices ($\alpha_{\rm r}>$0.4). 
For the five sources not detected
at 21 cm the upper limits in $\alpha_{\rm r}$ show that two have a flat, inverted
spectrum ($\alpha_{\rm r} < -$0.06 and $\alpha_{\rm r} < $ 0.21), while the other three 
can have both steep and
flat spectra.  The two multiple radio sources 
(LOCK\_6cm J105148+573248T and LOCK\_6cm J105237+573104T) 
and the sources LOCK\_6cm J105255+571950 and 
LOCK\_6cm J105347+573349
can be classified as Type II Fanaroff-Riley (FRII) 
radio galaxies  on the basis 
of their morphology and their radio luminosity (greater than 10$^{24.9}$ W/Hz, 
see last column of  Table~\ref{TableRX}), while all the other 9  
radio/X-ray  associations have radio luminosities typical of FRI radio 
galaxies (L$_{\rm 5~GHz}<$10$^{24.0}$ W/Hz).

\section{Discussion} 

\subsection{Radio spectral index} 

The spectral indices $\alpha_{\rm r}$ determined between 20 and 6 cm are listed in 
Table 3. Our calculation assumes that the radio sources have not varied 
significantly in the 8 years interval between the 20 cm 
observations (December 1990) and 
the 6 cm observations (January 1999) reported here. As shown by Oort 
\& Windhorst 1985 and by Windhorst et al. (1985), a significant variability 
occurs in only about 5\% of the mJy and sub-mJy radio sources population. 
Hence, the statistical analysis of the spectral index presented below 
should not be seriously biased  by source variability.  In our analysis 
we included both the measured values and the upper limits to $\alpha_{\rm r}$ 
given in Table 3 and used the software package ASURV which implements the 
methods described by Feigelson \& Nelson (1985) and Isobe et al. (1986).

\begin{figure} 
\hfill \psfig{figure=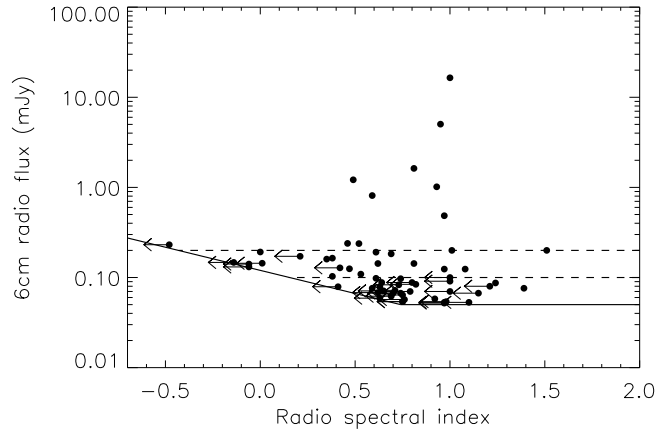,width=9cm} 
\caption[]{The 6cm flux as a function of the radio spectral index. The solid 
line shows the  minimum value of the 
spectral index below which no 20 cm detection is possible due to the 
flux density limit of the 20 cm survey.
The two dashed lines are at S$_{\rm 6~cm}$=0.1 mJy and at S$_{\rm 6~cm}$=0.2 mJy. 
}
\label{F11}
\end{figure}

We first studied the dependence of the spectral index distribution on the 
6 cm radio flux. However, because of the higher flux limit at 20 cm, 
if we consider all the 6 cm radio sources, the observed distribution 
of the spectral indices is a biased estimate of the true distribution. 
The main reason for this is that, given a minimum value for the 20 cm flux, 
for each value of the 6 cm flux there exists a minimum value of the 
spectral index below which no 20 cm detection is possible. 
In Fig. ~\ref{F11} 
we plot the 6 cm flux  as a function of the spectral indices 
(or their upper limits).  For illustrative purpose we show 
(solid line) the minimum 
spectral index, as a function of the 6 cm flux,  for sources which could 
have been detected in the more sensitive, central part of the 20 cm 
image (S$_{\rm lim}$=0.120 mJy).  The two dashed lines in 
 Fig. ~\ref{F11} are drawn at S$_{\rm 6~cm}$=0.1 mJy and 
at S$_{\rm 6~cm}$=0.2 mJy. 
As clearly shown in Fig. ~\ref{F11}, below 0.1 mJy 
there is  a bias against detection of sources with a radio 
spectral index flatter than 
$\sim$0.5. In fact, among the 32 sources  with 
S$_{\rm 6~cm}<$0.10 mJy, only 8 are detected at 20 cm. The situation is 
much better at S$_{\rm 6~cm}>$0.10 mJy, where we have 31 sources 
 with only 8 upper limits and a more relaxed bias (present only 
between 0.1 and 0.2 mJy, see Fig. ~\ref{F11})
against flat spectra sources. 
We have therefore considered in this
statistical analysis  only sources with S$_{\rm 6~cm}\geq$0.10 mJy. 

These sources have been divided in two
flux bins  (0.1$\leq$S$_{\rm 6~cm}<$0.2 mJy and S$_{\rm 6~cm}>$0.20 mJy, 
see dashed lines in Fig. ~\ref{F11}) 
and for each bin we
have computed, taking into account also the upper limits, the median
spectral index $\alpha_{\rm r\_med}$, the mean of the spectral index distribution 
$<\alpha_{\rm r}>$ and the fraction of flat and inverted spectrum sources
$f(\alpha_{\rm r}<0.5)$. The results, reported in Table~\ref{tab_alpha_r},
suggest the presence of a flattening of 
the radio spectral index in the
faint flux bin. The statistical significance of the difference between the
spectral index distributions in the two flux interval bins has been tested
using the statistical tests 
in the ASURV software package.
All tests suggest that the two distributions are different with a confidence
level of the order of 99.8\% (corresponding to about 3 $\sigma$), 
with mean and median spectral index 
flattening from $\sim$0.75 for  S$_{\rm 6~cm}\geq$ 0.2 mJy to $\sim$0.35 for 
0.1$\leq$S$_{\rm 6~cm}<$0.2 mJy. 

   \begin{table}
      \caption[]{The radio spectral index  versus 6 cm radio flux}
         \label{tab_alpha_r}
\begin{tabular}{cccc} 
&&& \\ \hline
4860 MHz Flux  & Median           & Mean           & $f(\alpha_{\rm r}<0.5)$  \\ 
Interval (mJy) & $\alpha_{\rm r\_med}$ & $<\alpha_{\rm r}>$   &                    \\ \hline
0.1$\leq$S$<$0.2~ & 0.37 $\pm$ 0.10  & 0.28 $\pm$0.08 & 11/19 $\sim$58\% \\
0.2$\leq$S$<$16.5 & 0.81 $\pm$ 0.14  & 0.73 $\pm$0.12 & ~3/12 $\sim$25\% \\ \hline
\end{tabular}
\end{table}

Also the fraction of flat spectrum sources appears to be different in the
two flux intervals, changing from $\sim$25\% for the brighter sample 
to $\sim$ 58\% for the fainter sample. 

\begin{center}
\begin{figure} 
 
\hfill \psfig{figure=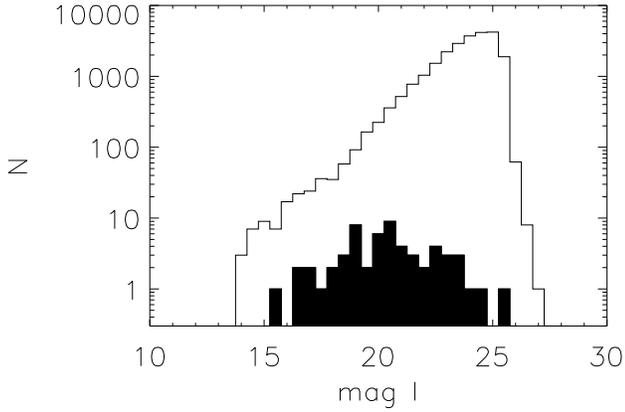,width=9cm} 

\caption[]{Magnitude distributions for the I band. 
The empty histogram shows the distribution of the whole optical 
sample, while the filled histogram shows the distribution 
of the optical counterparts of the 6 cm radio sources.  }

\label{F12}
\end{figure}   
\end{center}

Our results for the sources with 0.1 $<$ S$_{\rm 6~cm}\leq$0.2 mJy are in
excellent agreement with previous results in the same flux interval
(Donnelly et al., 1987) and with those found at even fainter fluxes
(Fomalont et al. (1991) for a sample selected at 5 GHz with 0.016
$<$ S$_{\rm 6~cm}\leq$0.2 mJy; Richards et al. (1999) for a sample 
selected at 8.5 Ghz with S$_{\rm med} \sim$ 0.03 mJy).

Vice versa, the suggestion from our data of a steeper average spectral index
for S$_{\rm 6~cm}\geq$0.2 mJy appears to be in contrast with the conclusion 
reached by Fomalont et al. (1991) and  Windhorst et al. (1993). 
They found, in fact, that while the median spectral index reaches 
the maximum steepness in the 10-100 mJy range (see Fig. 5 in 
 Windhorst et al., 1993), it decreases to $\sim$0.4 below 1 mJy and 
remains constant over almost a factor of 100 in flux density down to the 
faintest observed level ($\sim$ 0.016 mJy).  
Both our sample and those listed by Fomalont et al.
(see their Table 9) in the flux range 0.2 $\leq S_{\rm 6~cm}\leq$2 mJy are
however very small; only much larger surveys at these fluxes could
better determine at which flux the change in the spectral properties
(from steep spectra at high flux to flat spectra at low flux) is 
actually occurring. 

Finally we calculated the median of the spectral 
index distribution of the entire 6 cm sample with  S$_{\rm 6~cm}\geq$ 0.1 mJy. 
The median value of the 6 cm selected sample is 0.52$\pm$0.13. 
This value is consistent with the general trend of a flattening of the mean 
radio spectral index with increasing selection frequency 
(Fomalont et al., 1991, Windhorst et al., 1993) and with the  
recent results of Richards (2000), who found 
that, while the average spectral index 
for the 1.4 GHz radio sample in the HDF is 0.85$\pm$0.2, for the 8.5 GHz
selected sample it is 0.4$\pm$0.1. This flattening of the spectral 
distribution is consistent with the idea that high frequency samples 
preferentially select sources in which the radio emission is dominated 
by a flat-spectrum nuclear component (either a nuclear starburst with a 
free-free emission or a synchrotron self-absorbed AGN) and/or sources 
in which the radio emission is produced by thermal radio emission from 
large-scale star formation (Windhorst et al., 1993; Richards et al., 2000).

\subsection{Magnitude distribution and colour-colour diagram} 

In absence of spectroscopic data, the magnitude and colour distributions
of the optical counterparts can be used to derive some hints on the nature
of faint radio sources. 
The I  magnitude distribution of the optical counterparts of the radio sources
is shown as filled histogram in Fig.~\ref{F12}. The empty
histogram shows the magnitude distribution of the whole data set over the
region covered by the radio catalogue (a circle of 10 $^{\prime}$ radius).
These counts are in good agreement with the source counts found in other
surveys (Pozzetti et al., 1998). 
It is clear from Fig.~\ref{F12} that the I magnitude
distribution of the optical counterparts of radio sources reaches a maximum
at magnitudes (I $\sim$ 20 -- 21) well above our limiting magnitude,
consistently with the fact that a large fraction of the radio sources is
optically identified (see Sect. 4.2).  A similar result has been 
obtained by Richards et al. (1999) in  the identification of the 
microjansky radio sources in the Hubble Deep Field region.
Eighty-four out of 111 radio 
sources  have been identified down to I=25 mag, with the bulk of the 
sample identified with relatively bright (I$\leq$22 mag) galaxies.

\begin{figure} 
\hfill \psfig{figure=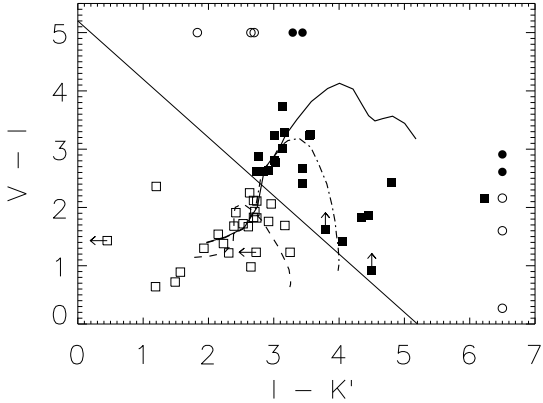,width=9cm} 
\caption[]{V-I versus I-K$^{\prime}$ colour for all the radio 
sources in the Lockman Hole. The diagonal line represents V-K=5.2. 
Typical evolutionary 
tracks from z=0 to z=2 are shown for early-type galaxies
(solid line: $\tau$=0.3 
IMF=Scalo age=12.5 Gyrs; dot dashed line : $\tau$=1)
and Sab Sbc   galaxies (dashed line: $\tau$=10 age=12.5 Gyrs;
courtesy of L. Pozzetti). 
Sources above the diagonal line are likely to be high redshift 
(z$\geq$0.5) early-type galaxies and are plotted as filled 
squares. Sources below the diagonal line can be late-type, star 
forming galaxies at all redshifts or  early-type galaxies at 
low redshift (z$\leq$0.5) and are plotted as empty squares.
In the upper and right part 
of the plot we show objects (plotted as circles) for which no 
information in V and K$^{\prime}$ magnitudes, respectively, 
are available. We tentatively assigned a classification on the basis 
of the available colour: filled circles are sources with V-I$>$2.5 or 
I-K$^{\prime}>$3.0 ($i.e.$ sources probably similar to the 
sources plotted as filled squares), open circles are sources with 
V-I$<$2.5 or I-K$^{\prime}<$3.0.

}
\label{F13}
\end{figure}

In Fig.~\ref{F13}  we show the  V-I versus 
I-K$^{\prime}$ colour for all the radio sources in the Lockman Hole with a
magnitude measurement in all three bands (42 objects) or an
upper limit either in V or in K$^{\prime}$ (4 objects; see symbols with
arrows in the figure). 
Typical evolutionary tracks from z=0 to z=2, computed convolving
the models' spectral energy distributions with the appropriate filters, 
are shown for elliptical galaxies (solid line: $\tau$=0.3 
IMF=Scalo age=12.5 Gyrs; dot dashed line : $\tau$=1)
and late--type (Sab--Sbc) galaxies (dashed line: $\tau$=10 age=12.5 Gyrs;
courtesy of L. Pozzetti). 

This comparison between data and models suggests that all (or most of)
the objects above the diagonal line (filled squares) are likely to be high
redshift (z $\ge$ 0.5) early--type, passively evolving galaxies. The other
objects (empty squares) can be late--type, star--forming galaxies at all
redshifts or early--type galaxies at low redshift (z $\le$ 0.5).  

In the upper and right part of the figure we show objects for which
no information in V and K$^{\prime}$ magnitudes, respectively, are
available. This is due to the fact that the fraction of the radio data image
covered by the V and K catalogues are $\sim$95\% and $\sim$70\%. These
objects, shown as circles, are tentatively assigned a classification
on the basis of the available colour (see figure caption). 

\subsection{Radio flux versus optical magnitude}  

\begin{figure} 
\hfill \psfig{figure=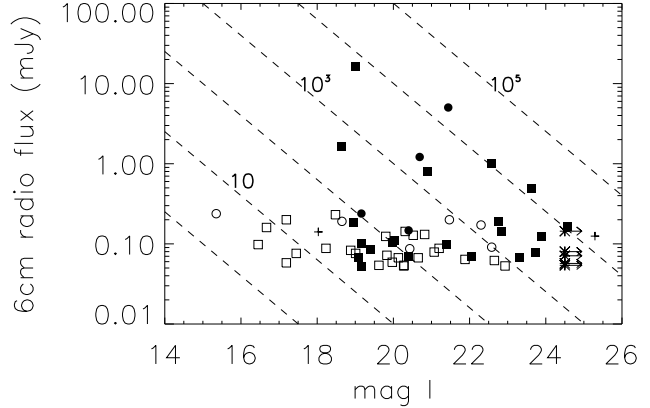,width=9cm}   
\caption[]{The I band magnitude 
versus the  6 cm radio flux for all the 63 radio sources. 
Symbols as in Fig.~\ref{F13} except for the two 
crosses that show the objects for which only an I band 
magnitude is available and the five arrows that show
the radio sources without identification in the V I 
and K$^{\prime}$ band. 
The lines 
represent different radio to optical ratios R, 
corresponding to R=1,10,10$^2$,10$^3$,10$^4$,10$^5$.} 

\label{F14}
\end{figure}

Fig.~\ref{F14} shows the  6 cm radio flux versus the I band magnitude 
for all the 63 radio sources (I band upper limits are shown for the 5
radio sources with no optical counterpart). Superimposed 
are the lines corresponding to constant values for the observed 
radio-to-optical ratios $\it R$, defined  as $\it R$ = S $\times$ 10$^{0.4(I-12.5)}$,
where S and I are the radio flux in mJy and the apparent magnitude of
sources respectively. The symbols are the same as those defined in
the previous figure, except for the crosses, which represent the
objects for which only an I band magnitude is available and therefore
were not shown in the colour -- colour diagram.

This figure,  although limited to a relatively narrow range of radio 
fluxes (see Fig. 8a in Kron et al. (1995) for a similar plot at 
higher flux level),   
shows that the two classes of objects defined on the
basis of the colour--colour diagram have a different distribution
in the radio flux--optical magnitude diagram. In particular,
all the objects at large radio-to-optical ratios ($\it R$ $\ge$ 1000) 
have colours typical of passively evolving galaxies at relatively
high redshift. On the basis of this plot we would conclude that
also the five radio sources without optical identification are likely to
belong to this class. No object of this class, instead, appears in this
plot at low values of $\it R$  ($\it R$ $\le$ 30), consistent with previous
findings (see, for example, Gruppioni et al., 1999) that objects
with these values of R are mainly identified with star--forming galaxies.
The situation is less well defined for intermediate values of $\it R$
( 30 $ < \it R < $ 1000) where the two populations defined on the basis
of Fig.~\ref{F13} are not clearly separated in this plot. 
Only spectroscopic
data can help in better defining the relative proportions of the
two populations in this range of $\it R$. In any case, this analysis
already allows us to conclude that at least about 50\% of the radio sources
in a 5 GHz selected sample with limiting flux S$_{\rm 6~cm}\geq$0.05 mJy
is associated to early--type galaxies. A similar conclusion was reached
by Gruppioni et al. (1999) for a 1.4 GHz selected sample
(S$_{\rm 21cm}\geq$0.2 mJy), for which a substantial fraction of spectroscopic
identifications was available.

\subsection{Extremely Red Radio Galaxies} 

\begin{figure} 

\hfill \psfig{figure=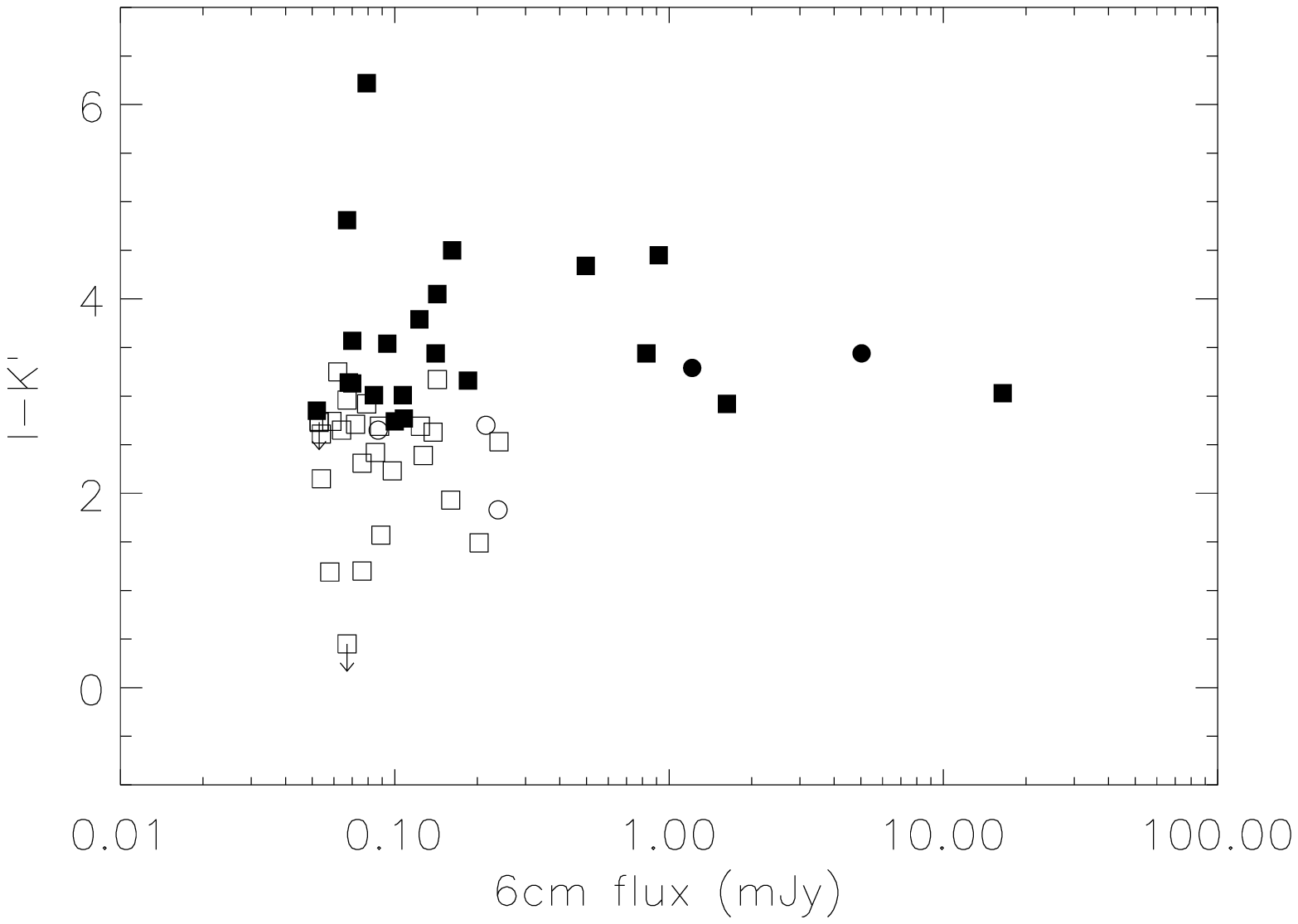,width=8.5cm} 

\hfill \psfig{figure=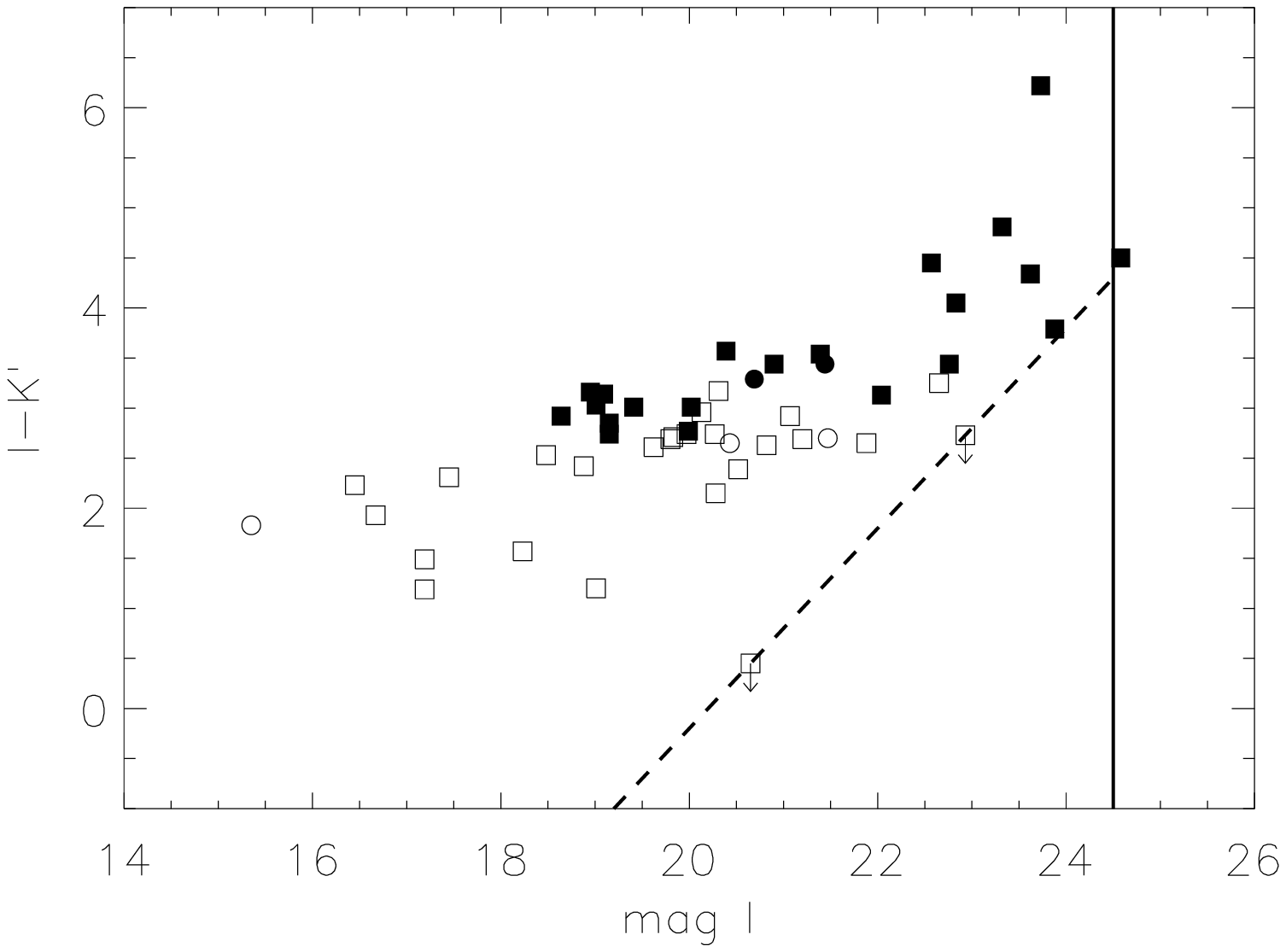,width=8.5cm} 

\hfill \psfig{figure=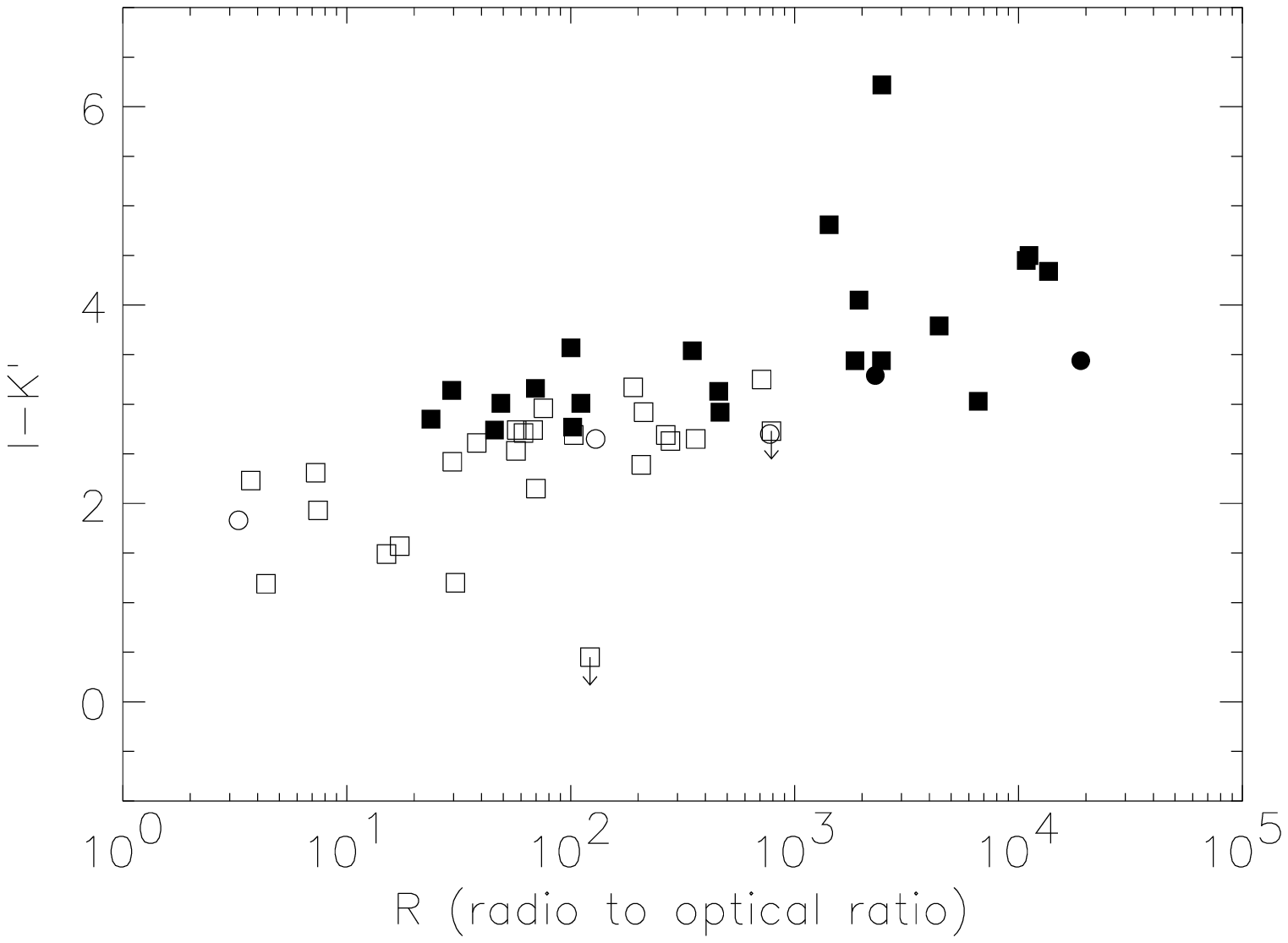,width=8.5cm} 

\caption[]{The   I-K$^{\prime}$ colour as function of the total radio 
flux (top panel), of the I band magnitude (central panel) and 
of the radio-to-optical ratio R (bottom panel). Symbols as in 
Fig.~\ref{F13}.  The vertical line in the central panel 
shows the magnitude limit I=24.5 mag of our optical data while 
the dashed slanted line shows the K$^{\prime}$=20.2 mag limit. 
Sources within the two lines can not be detected due to the 
magnitude limits.  }

\label{F15}
\end{figure}

In Fig.~\ref{F15} we show the I-K$^{\prime}$ colour as a function 
of the radio flux, of the I magnitude and of the radio-to-optical ratio 
$\it R$.  While no obvious correlation is seen between I-K$^{\prime}$ and 
radio flux,  there appear to be significant correlations between
I-K$^{\prime}$ and both I magnitude and radio-to-optical ratio $\it R$.

Both of them are significant at more than 5$\sigma$ level on the basis 
of the  Spearman rank test. 
A similar trend for the optically fainter radio sources to have redder
I-K colours was found by Richards et al. (1999)
in their identification of fainter radio sources in the HDF and SSA13
fields. 
From  Fig.~\ref{F15}   it is evident that almost all the counterparts 
of the radio sources in the magnitude range 22.5$<$I$<$24.5 
or with a radio-to-optical
ratio greater  than 1000 are red objects with  I-K$^{\prime}>$ 3 and that 
a  high fraction (6/10) of these faint objects can 
be classified as Extremely Red Objects (EROs) 
on the basis of a colour  I-K$^{\prime}$  $>$ 4 (McCarthy et al., 1992; Hu \& 
Ridgway, 1994). 
These EROs sources are not present at magnitudes I brighter than 
I$\sim$22.5 mag and among sources with a radio-to-optical ratio 
lower than 1000. 

The very red colours of EROs are well known to be consistent with
both old passively evolving distant (z $>$ 0.8) elliptical galaxies
(e.g. Cohen et al., 1999; Spinrad et al., 1997) and dust--reddened
starburst galaxies (e.g. Cimatti et al., 1998; Smail et al., 1999).
The identification of the nature of EROs is an important test for
models of galaxy formation and evolution and therefore  
the relative proportions of these two classes of objects among EROs
has been lively debated in the last few years. However, because of the
faintness of these objects in the optical bands, very few spectroscopic
confirmations were available until recently. Various authors, in absence
of optical spectroscopy, used  colour-colour diagrams and the overall
spectral energy distribution  to attempt to discriminate 
between passively evolving elliptical or dusty star--forming galaxies 
(Pozzetti \& Mannucci, 2000; Willott et al., 2001). 
Very recently, Cimatti et al. (2002), on the basis of extensive VLT
spectroscopy of a complete sample of EROs with K $<$ 20 have convincingly
shown that the EROs population is indeed made up of both 
passively evolving elliptical and dusty star--forming galaxies
and that the two classes of objects appear to be about equally populated
and cover a similar redshift range (0.7 $<$ z $<$ 1.5). The average
spectrum of the star--forming EROs suggests a substantial dust 
extinction with $E(B-V)\geq$0.5.

Not yet having spectroscopic data for these radio--selected EROs, we here  try
to set some constraints on their nature using information from our own radio
data and from existing ISO data. If these sources were dusty starbursts
at z $\sim$ 1, from the average dust extinction of $E(B-V)=$0.5
suggested by Cimatti et al. (2002) for this class of objects we estimated 
$\Delta$I $\sim$ 2.7 mag and $\Delta$(I$-$K$^{\prime}$) $\sim$ 1.9 
using the extinction law of Calzetti (1997) for objects at z$\sim$1.  
A  dust extinction of $E(B-V)=$0.5 is in agreement with the 
extinction estimated by Willott et al. (2001) 
studying six extremely red sources ($R-K>$5.5) from the 7C radio sample.
From the fit of their spectral energy distribution they found 
six best-fit galaxy models with a dust extinction in the range 
0.2$\leq E(B-V) \leq $0.8. 

Using the above extinction  corrections the  
intrinsic colours and the radio to optical ratios of the 6 EROs here selected
(2.15 $<$ I$-$K$^{\prime}$ $<$ 4.32 and 50 $<$  radio to optical ratio $<$ 1200)
would become formally consistent with those typical of late type galaxies. 
However, under this assumption, their star formation rate (SFR) in massive
stars, estimated from the radio emission using
SFR($M\geq5M_{\odot})=L_{1.4}/(4.0\times10^{21} W Hz^{-1})$ (Cram et al., 1998)
and assuming z $\sim$ 1, would be in the range 100-2300 $M_{\odot}$ yr$^{-1}$.
Because of the well established correlation between SFR and far infrared
emission (SFR$(M\geq5M_{\odot})=L_{60_{\mu m}}/(5.1\times10^{23} W Hz^{-1})$;
Cram et al., 1998), and assuming the spectral energy distribution of M82,
these values of SFR would imply 15 $\mu$m fluxes in the range 3-70 mJy and 
90 $\mu$m fluxes in the range 10-200 mJy. However ISO observations
of this field show that none of the six radio selected EROs selected is 
detected at 90 $\mu$m down to a flux limit of $\sim$ 60 mJy (Rodighiero et al.,
in preparation) and only two of them (LOCK\_6cm\_J105255+571950 and 
LOCK\_6cm\_J105314+573020) are detected at 15 $\mu$m with a flux of 
1.5 and 0.7 mJy respectively, with an upper limit of $\sim$ 0.30 mJy 
for the other four sources (Fadda et al., 2002; Fadda et al. in preparation). 

From this comparison of optical, radio and ISO data we conclude that most
of our radio--selected EROs sources are likely not to be associated
with dusty starburst galaxies, but rather with early-type galaxies, hosting
an Active Galactic Nucleus (AGN) responsible of the radio activity, 
although some  contribution from the radio emission also from a 
nuclear starburst activity can not be excluded. 

This conclusion is also in agreement with most of previous spectroscopic
findings on the nature of extremely red radio sources. Laing et al. (1983) 
showed that the light from the reddest z $\sim$ 1 radio
galaxies in the 3CRR sample (3C65) is dominated by an old ($\sim$4Gyr)
stellar population. Subsequently, two  extremely red radio galaxies at
z$\sim$1.5 have been discovered in the follow up of the faint Leiden
Berkeley Deep Survey (Dunlop et al., 1996; Dunlop, 1999). Keck spectroscopy of
these galaxies shows that also in this case their red colours are due to an
old stellar population ($\geq$3 Gyr) and not to reddening by dust. A similar
result has been recently obtained by Willott et al. (2001)
studying six extremely red sources ($R-K>$5.5) from the 7C radio sample. The
only example so far of an  extremely red object  with I$-$K$>$4  
selected from a radio
survey and identified with a dusty star--forming galaxy has been found by 
Afonso et al. (2001) studying 
optical counterparts of the radio sources from the Phoenix Deep Radio Survey
(see also Waddington et al. (1999) for a similar dusty galaxy with 
I$_{814}$-K=2.0 and a likely redshift of z=4.424). 
 
Finally, it is interesting to note that the six radio selected EROs represent 
only $\sim$ 2\% of all the sources with I-K$^{\prime}>$4 present in the area 
covered by the K$^{\prime}$ data ($\sim$ 300 sources with I-K$^{\prime}>$4 
over $\sim$ 220 arcmin$^2$). This fraction could approximately double if
most of the five radio sources with no I band counterpart also belong to
the EROs class, as suggested by the observed trend between colour and
I magnitude (see Fig. ~\ref{F15}).
Assuming that their radio luminosity is due to 
AGN activity, the small fraction of radio detection suggests a relatively
small AGN content in the optically/near-infrared selected EROs population. 
This result is qualitatively in agreement with what suggested by the recent
deep X-ray surveys. In the Hubble Deep Field North Caltech area, only 4 of
the 33 EROs in the field  have a hard X-ray detection in the $\sim$ 220 ksec
exposure with the Chandra satellite (Hornschemeier et al., 2001).
However, it is well known that luminous radio sources have lifetimes
much smaller than the age of the early galaxies hosting these AGNs
($\geq$3-4 Gyr). Therefore, if individual radio sources have lifetimes of 
about $\sim10^8$ years (Condon, 1992), then the number of EROs undergoing 
radio activity during their life would be much higher than that observed. 

\section{Conclusion}

We have used the VLA radio telescope to image at 6 cm a circular region  
(10 arcmin radius) to a 4.5 $\sigma$ limit of $\sim50\mu$Jy in the Lockman 
Hole. The region is centered around the ROSAT ultra-deep HRI field. 
A complete sample of 63 radio sources has been obtained. The differential 
source counts are in good agreement with those obtained by other surveys,
confirming that while the slope of the counts shows a 
flattening below a flux density of $\sim$1-3 mJy, 
there is no indication for a significant change in the slope 
in the 6 cm counts between $\sim$0.1 and 1 mJy. 

The availability of a 20 cm survey down to $\sim$0.12 mJy in the same 
area (de Ruiter et al., 1997), gave us the opportunity to study the radio 
spectral index as function of the radio flux. Dividing our 6 cm sample in 
two subsamples with fluxes lower and greater than 0.2 mJy, 
we find a flattening of the radio spectral index 
and an increase of the population 
of flat spectra radio sources in the fainter flux bin.
Several explanations for the observed 
flattening compared with sources detected at higher flux density level are 
possible, including an increasing number of synchrotron
self-absorbed AGNs among the microjansky population, and/or a rising 
component of thermal radiation from active star formation. 

Using the available optical images in V and I bands, we performed
the optical identification of the radio sources using the Likelihood Ratio
Analysis.  We found a likely optical identification for 58 of the 63 radio
sources. From the reliability associated
to the likelihood analysis we estimate that at most four of the proposed
identifications may be spurious and therefore the identification rate with the
``correct'' optical counterpart is in the range $\sim$86\% - 92\%. 
Moreover, for a subsample of 51 radio sources we have 
data also in the K$^{\prime}$ band. For this subsample we find  K$^{\prime}$
counterparts for 49 sources (96\% of identifications). 
These high identification rates are consistent with the observed
magnitude distribution of the optical counterparts, which appears
to reach a maximum at magnitudes (I $\sim$ 20 - 21) well above
the limiting magnitude of our optical data.

A cross correlation between the radio catalogue and the ROSAT 
catalogue and the XMM-Newton 
source list (106 sources detected in the 0.5-2.0 keV band down to 
a flux limit of 3.8$\times$ 10$^{-16}$ erg cm$^{-2}$ s$^{-1}$, 
Lehmann et al., 2002) gives 13 reliable radio/X-ray associations. 
The percentages of the radio/X-ray associations are  
$\sim$ 21 per cent of the radio sample and $\sim$ 12 per cent of 
the XMM/X-ray sample. 

From an analysis of a colour--colour diagram (V-I versus I-K$^{\prime}$)
we divided the optical counterparts in two classes. The objects in the first
class are likely to be high redshift (z $\ge$ 0.5) early--type, passively
evolving galaxies, while the objects in the second class can be both
late--type, star--forming galaxies at all redshifts and 
early--type galaxies at low redshift (z $\le$ 0.5). This
separation in two classes appears to have a physical basis, 
related to different radio
emission mechanisms, as further supported by the fact
these two classes of objects have a quite different distribution
in the radio flux--optical magnitude diagram. In particular,
all the objects at large radio-to-optical ratios ($\it R$ $\ge$ 1000) 
have colours typical of passively evolving galaxies at relatively
high redshift. On the basis of this plot we suggest that
also the five radio sources without optical identification are likely to
belong to this class. No object of this class, instead, appears in this
plot at low values of $\it R$  ($\it R$ $\le$ 30), consistent with previous
findings (see, for example, Gruppioni et al., 1999) that objects
with these values of R are mainly identified with star--forming galaxies.
From this analysis based on colours and the radio-to-optical ratios 
we conclude that at least 50\% of the radio sources
in a 5 GHz selected sample with limiting flux S$_{\rm 6~cm}\geq$0.05 mJy
is associated to early--type galaxies. A similar conclusion was reached
by Gruppioni et al. (1999) for a 1.4 GHz selected sample
(S$_{\rm 21cm}\geq$0.2 mJy), for which a substantial fraction of spectroscopic
identifications was available.

Our data also show a significant correlation between 
I-K$^{\prime}$ colour and both I magnitude and radio-to-optical ratio $\it R$,
with the redder galaxies being associated with optically fainter radio 
sources and with higher radio-to-optical ratio. In particular, 
almost all the counterparts of the radio sources  in the magnitude range 
22.5$<$I$<$24.5 are red objects with  I-K$^{\prime}$  $>$ 3 and a high 
fraction (6/10) of these red objects are EROs with I-K$^{\prime}$  $>$ 4. 
By combining our radio data with existing ISO data we conclude
that most of our six radio--selected EROs sources are likely not to be
associated with dusty starburst galaxies, but rather with early-type galaxies,
hosting an AGN responsible of the radio activity. 

The six radio selected EROs represent only $\sim$ 2\% of the 
optically selected EROs present in the field. If their radio luminosity is 
indeed a sign of AGN activity, the small fraction of radio detections suggests
that the optically/near-infrared selected EROs population contains
a relatively small fraction of active AGN, in agreement with what suggested
by recent deep X-ray surveys (Hornschemeier et al., 2001).

The very high percentage of optical identifications and the availability 
of near infrared and deep X-ray data 
make this sample one of the best samples of radio selected sources available 
at $\mu$Jy flux density level and well suited to study in detail the nature 
of faint 6 cm radio sources.  With this aim, 
we already obtained spectroscopic data for about one third of the 
radio sample, while a programme to obtain spectroscopic 
data for all the sources  is ongoing.  
The results of the optical spectroscopic  classification and a more detailed
comparison with the XMM source list will be presented in a forthcoming paper.


\begin{acknowledgements}

This work was supported by the Italian Ministry for University 
and Research (MURST) under grant COFIN01 and by Italian Space Agency 
(ASI). We thank L. Pozzetti for providing us the  evolutionary 
tracks of Fig.~\ref{F13}, D. Fadda and G. Rodighiero 
for providing us ISOCAM and ISOPHOT~ data before publication, and the 
referee  R.A. ~Windhorst for useful comments. 
G.P.S. acknowledges support under DLR grant 50 OR 9908. 

\end{acknowledgements}

\end{document}